\documentclass[a4paper]{article}
\usepackage{hyperref}
\usepackage{makeidx}
\makeindex
\usepackage{latexsym}
\usepackage{graphics}
\usepackage{epsfig}
\usepackage{amsmath}
\usepackage{amssymb}
\usepackage{fancybox}
\usepackage{bm}
\usepackage{amsthm}
\usepackage{mathrsfs}
\usepackage{fancyhdr}
\usepackage{ascmac}

\def\stars{{\star\star}}
\def\bcdot{{\circ}}
\def\bcdots{{\bcdot\bcdot}}
\def\bcdott{{\bcdots\bcdots}}

\def\TM{{T\hspace{-.2em}\M}}
\def\TsM{{T^*\hspace{-.2em}\M}}
\def\TMM{{T\hspace{-.2em}\MM}}
\def\TsMM{{T^*\hspace{-.2em}\MM}}

\def\aaa{{\mathfrak a}}
\def\AAA{{\mathfrak A}}

\def\bbb{\mathfrak{b}}
\def\BBB{\mathfrak{B}}

\def\ooo{\mathfrak{o}}

\def\C{\mathbb{C}}
\def\Z{\mathbb{Z}}

\def\uuu{\mathfrak{u}}

\def\TTT{{\mathfrak T}}

\def\td{d_\www}

\def\vvv{\mathfrak{v}}
\def\VVV{\mathfrak{V}}

\def\FFF{\mathfrak{F}}

\def\GGG{\mathfrak{G}}
\def\ggg{\mathfrak{g}}

\def\sss{\mathfrak{s}}

\def\SSS{{\mathfrak S}}

\def\LL{{\cal L}}

\def\LLL{{\mathfrak L}}

\def\HH{{\cal H}}

\def\HHH{{\mathfrak H}}

\def\BBB{{\mathfrak B}}
\def\bbb{{\mathfrak b}}

\def\C{\mathbb{C}}
\def\CCC{{\mathfrak C}}
\def\ccc{{\mathfrak c}}

\def\Varepsilon{{\mathcal E}}

\def\eee{{\mathfrak e}}

\def\M{{\cal M}}
\def\MM{\mathscr{M}}
\def\MMM{\mathfrak{M}}

\def\N{\mathbb{N}}

\def\R{\mathbb{R}}

\def\RRR{\mathfrak{R}}

\def\vomega{{\mathfrak w}}
\def\www{{\mathfrak w}}
\def\WWW{{\mathfrak W}}

\def\tSSS{\widetilde{\SSS}}
\def\twww{\widetilde{\www}}

\def\sc{{\sss\hspace{-.1em}\ccc}}

\def\MMvac{\M_{\hspace{-.1em}vac}}
\def\LLvac{{\LL_{\hspace{-.1em}vac}}}
\def\LLschw{\LL_{\hspace{-.1em}\mathrm{Schw}}}

\def\Varepsilonvac{{\Varepsilon}_{\hspace{-.1em}vac}}

\def\Mvac{M_{\hspace{-.1em}vac}}
\def\WWWgr{\WWW_{\hspace{-.1em}G\hspace{-.1em}R}}
\def\Mgr{M_{\hspace{-.1em}G\hspace{-.1em}R}}
\def\sssvac{\sss_{\hspace{-.1em}vac}}
\def\sssgr{\sss_{\hspace{-.1em}G\hspace{-.1em}R}}
\def\Sgr{S_{\hspace{-.1em}G\hspace{-.1em}R}}

\def\sc{{\sss\hspace{-.1em}\ccc}}
\def\scgr{{\sss\hspace{-.1em}\ccc_{\hspace{-.1em}G\hspace{-.1em}R}}}
\def\scvac{\sss\hspace{-.1em}\ccc_{\hspace{-.1em}vac}}
\def\scgr{\sc\hspace{-.1em}\ccc_{\hspace{-.1em}G\hspace{-.1em}R}}
\def\sc{{\sss\hspace{-.1em}\ccc}}

\def\mugr{{\mu_{\hspace{-.1em}G\hspace{-.1em}R}}}
\def\nugr{{\nu_{\hspace{-.1em}G\hspace{-.1em}R}}}

\def\eeevac{{\eee_{\hspace{-.1em}vac}}}
\def\wwwvac{{\www_{\hspace{-.1em}vac}}}
\def\RRRvac{{\RRR_{\hspace{-.1em}vac}}}
\def\TTTvac{{\TTT_{\hspace{-.1em}vac}}}

\def\FF{{\mathcal F}}
\def\cP{{\hspace{-.1em}c\hspace{-.05em}\mathrm{P}}}
\def\Tr{\mathrm{Tr}}
\def\det{\mathrm{det}}
\def\cP{{\hspace{-.1em}c\hspace{-.05em}\mathrm{P}}}
\newtheorem{definition}{Definition}[section]
\newtheorem{theorem}[definition]{Theorem}

\newtheorem{example}[definition]{Example}

\newtheorem{remark}[definition]{Remark}
\newtheorem{Pstat}[definition]{Physical statement}
\newcommand{\QED}{\nobreak \ifvmode \relax \else
      \ifdim\lastskip<1.5em \hskip-\lastskip
      \hskip1.5em plus0em minus0.5em \fi \nobreak
      \vrule height0.75em width0.5em depth0.25em\fi}

\begin{document}
\title{
Symplectic structure for general relativity and Einstein--Brillouin--Keller quantization
}
\author{Yoshimasa Kurihara\footnote{yoshimasa.kurihara@kek.jp}
\\
{\it\footnotesize The High Energy Accelerator Organization (KEK), 
Tsukuba, Ibaraki 305-0801, Japan}
}
\date{}
\maketitle

\begin{abstract}
The Hamiltonian system of general relativity and its quantization without any matter or gauge fields are discussed on the basis of the symplectic geometrical theory.
A symplectic geometry of classical general relativity is constructed using a generalized phase space for pure gravity.
Prequantization of the symplectic manifold is performed according to the standard procedure of geometrical quantization.
Quantum vacuum solutions are chosen from among the classical solutions under the Einstein--Brillouin--Keller quantization condition.
A topological correction of quantum solutions, namely the Maslov index, is realized using a prequantization bundle.
In addition, a possible mass spectrum of Schwarzschild black holes is discussed.
\end{abstract}

\section{Introduction}
In early studies of quantum mechanics, the Bohr--Sommerfeld quantum condition played an important role in the selection of quantum states of classical systems with periodic motion.
Quantum theories appearing before the establishment of modern quantum mechanics are referred to as the old quantum mechanics.
They not only provide calculation tools for microscopic systems but also assist physicists in finding/understanding the entity of quantum mechanics.
The Bohr--Sommerfeld quantum condition was initially applied to a single-body problem such as a hydrogen atom.
For a multi-body problem, this method can be independently applied to each variable after variable separation\cite{epstein1916,epstein1919}, if possible. 
Further extension of this method to a non-variable-separation type of problem was proposed by Einstein\cite{Einstein:422242} and Brillouin\cite{brillouin1926remarques}, and an essential improvement  was presented by Keller\cite{KELLER1958180,KELLER196024}.
This method is now referred to as the Einstein--Brillouin--Keller (EBK) quantization method.
The EBK quantization method was further refined by mathematicians using symplectic geometry\cite{maslov75theory,Arnol'd1967}, in which a topological index, which was first introduced owing to physical considerations by Keller, is defined as a generator of the first cohomology of the stable Lagrangian-Grassmanian with integer coefficients (see, for example, Ref.\cite{bates1997lectures}). 
The EBK quantization method remains under investigation to date.
For example, the relationship between the EBK method and the Heisenberg matrix mechanics has been discussed in Ref.\cite{PhysRevA.54.1820} and some elucidative examples of the method have been provided in Ref.\cite{2004AmJPh..72.1521C}.
The EBK quantization method has also been used to analyze quantum chaos\cite{gutzwiller1991chaos,2005Chaos..15c3107D}.
In this paper, we propose a new application of this method to a space-time manifold under general relativity.

We focus on the applications of the EBK quantization method to general relativity.
The EBK quantization method requires discretization of the action integral represented using symplectic forms, and the ground state of a discretized system is given according to the topological structure of an integration region, i.e. the so-called Maslov index\cite{maslov75theory,maslov1965wkb}.
First, general relativity is cast into the symplectic formalism for the application of EBK quantization.
The symplectic formalism of general relativity has been discussed by many authors\cite{Epp:1995uc,Peres1962,carter1968,PhysRev.170.1195,PhysRev.177.1929,hojman1973,ashtekar1982,0264-9381-8-11-016,10.2307/52025,0264-9381-9-8-015,PhysRevD.49.2872,doi:10.1063/1.1489501,Kocherlakota:2019frz}.
In contrast to, for example, the method proposed by Ashteka\cite{ashtekar1982,ashtekar1991lectures}, our formalism does not employ a $(3\hspace{-.em}+\hspace{-.em}1)$-decomposition and treats all space time directions equally. 
Therefore, the $GL(4)$ symmetry of the theory is apparent in our method.
The Maslov index appearing in the EBK quantization of general relativity is identified as a characteristic class (the second Chern class) of a space time manifold given as a solution to the Einstein equation.
This characteristic class is induced by a principal bundle with the co-Poincar\'{e} group\cite{doi:10.1063/1.4990708} as a structural group.  
The co-Poincar\'{e} group is a novel symmetry introduced in a previous work by the author.
In the present study, the integral value of the Maslov index is ensured using the Chern--Weil theory\cite{zbMATH03077491, frankel_2011, roe1999elliptic}.

In this study, the EBK quantization of general relativity is developed as follows.
First, Section 2 introduces a completely $GL(4)$-covariant Hamiltonian formalism of general relativity in a four-dimensional space-time manifold using a method similar to the de\hspace{.1em}Donder--Weyl Hamiltonian theory\cite{10.2307/1968645, donder1930theorie, KASTRUP19831}.
In addition, a canonical pair of phase-space variables is determined owing to the introduced Hamiltonian formalism.
Section 3 proves the main theorem of this study ({\bf Theorem \ref{sypform}}), which states that the above-mentioned canonical pair has a symplectic structure and forms a symplectic manifold.
The first objective of this study is to prove {\bf Theorem \ref{sypform}}.
Owing to the symplectic structure of general relativity provided here, a standard method of geometrical quantization can be applied to the canonical pair of general relativity.

Next, Section 4 discusses the geometrical quantization of general relativity.
According to the standard procedure of geometrical quantization, a Legendre submanifold and a prequantization bundle are introduced.
In particular, a contact manifold is introduced using a symplectic manifold immersed in a five-dimensional manifold; then, a Legendre submanifold is obtained as a submanifold of the contact manifold restricted by vacuum solutions of the Einstein equation.
Further, a prequantization manifold is introduced on the Legendre submanifold.
Prequantization is the process of constructing an $n$-dimensional Hilbert space on an $2n$-dimensional symplectic manifold; this reduction of dimensionality is called a ``polarization''.
The last step of geometrical quantization is to construct a Hermitian operator on the above-mentioned Hilbert space.
The last step is not considered in this study; instead, EBK quantization is introduced in Section 5.

EBK quantization is a method that gives approximated quantum states from the classical solutions of a given system.
The second and main objective of this study is to establish a mathematically rigorous and computable EBK quantization condition of general relativity.
The quantization condition proposed herein can be considered as a type of physical conjecture, because a complete quantum gravitation theory is not known to exist at present; hence, it is possible to immediately judge whether this quantization condition chooses the correct quantum state.
In the future, this conjecture should be evaluated through experiments or observations.   
In addition, an EBK quantization criterion is applied to a vacuum solution of the classical Einstein equation, and an energy spectrum and the ground state of a Schwarzschild black hole are extracted in Section 5.
Finally, Section 6 summarizes geometrical quantization and its application to general relativity.

Note that this study discusses only pure gravity without the cosmological constant or gauge/matter fields.

\section{Preliminaries}
This section presents the terminologies and notations of differential geometry and classical general relativity used in the present study.
We primarily (but not completely) follow the terminologies and notations employed in Refs.\cite{darling_1994, fre2012gravity}.

%
%
\subsection{Classical general relativity}\label{cgr} 
A four-dimensional smooth pseudo-Riemannian manifold $(\MM,\bm g)$ is considered, where $\MM$ is a manifold with $GL(4)$ and $\bm g$ is a metric tensor in $\MM$.
The determinant of the metric tensor is assumed to be negative.
A local frame field in an open neighbourhood $U_p\subset\MM$ at $p\in U_p$ is represented as $x^\mu(p)$. 
Orthonormal bases in $T_{\hspace{-.1em}p}\MM$ and $T^*_{\hspace{-.1em}p}\MM$ are introduced as $\partial_\mu$ and $dx^\mu(p)$, respectively.
The abbreviation $\partial_\mu:=\partial/\partial x^\mu(p)$ is used throughout this study.
Spaces of rank-$q$ tensors ($q=0,1,2,\cdots$) on trivial tangent space $\TMM:=\bigcup_p T_{\hspace{-.1em}p}\MM$ and cotangent space $\TsMM:=\bigcup_p T^*_{\hspace{-.1em}p}\MM$ are denoted as $V^q(\TMM)$ and  $\Omega^q(\TsMM)$, respectively.
Bases associated with trivial tangent and cotangent bundles, $\partial_\mu$ and  $dx^\mu$ respectively, are referred to as the trivial bases hereafter.
Riemannian manifold $(\MM,\bm g)$ is referred to as a global manifold.

Local manifold $(\M_p,\bm\eta)$ with Poincar\'{e} symmetry $ISO(1,3)=SO(1,3)\ltimes T^4$ can be found at any point $p\in\MM$, where $T^4$ is a four-dimensional translation group.
This expression is known as the Einstein equivalence principle in physics, and $\M_p$ is identified as the local inertial system in physics.
Trivial bundle $\M:=\bigcup_p\M_p$ is considered as the principal bundle of $\MM$ with a Poincar\'{e} group as a structural group.
Metric tensor $diag[{\bm \eta}]=(1,-1,-1,-1)$ is exploited in this study.
In addition, tangent space $\TM:=\bigcup_p T_{\hspace{-.1em}p}\M$ and cotangent spaces $\TsM:=\bigcup_p T^*_{\hspace{-.1em}p}\M$ are introduced.
A map $\Varepsilon:(\MM,\bm g)\rightarrow(\M,\bm \eta):{\bm g}\mapsto{\bm \eta}$ is represented using a trivial basis on any chart of $\MM$ as
\begin{eqnarray*}
\eta^{ab}&=&
\Varepsilon^a_{\mu_1}(p)\Varepsilon^b_{\mu_2}(p)g^{\mu_1\mu_2}(p),
\end{eqnarray*}
where $\eta^{ab}=[{\bm \eta}]^{ab}$ and $g^{\mu\nu}=[{\bm g}]^{\mu\nu}$ are the metric tensors on $\M$ and $\MM$, respectively.
The Einstein convention for repeated indices is used throughout this study.
Function $\Varepsilon^a_{\mu}(p)=[{\bm \Varepsilon}]^a_{\mu}(p)$ is referred to as the vierbein.
In our representation, base manifolds of tensors are distinguished by component indices, i.e. Greek suffixes in $\MM$ and Roman suffixes in $\M$.

Owing to homomorphism $\TsMM\cong \TsM$, differential form $\aaa\in\Omega^q(\TsM)$ and its pull-back  $\Varepsilon^\sharp\hspace{-.1em}\aaa\in\Omega^q(\TsMM)$ are equated to each other and simply denoted as $\aaa\in\Omega^q$\footnote{Differential forms are represented using Fraktur letters such as ``$\AAA,\aaa,\BBB,\bbb,\CCC,\ccc,\cdots$'', and a pull-back with respect to a map $\bullet$ is indicated as $\bullet^\sharp$ in this study.}.
Orthonormal base vectors in $\TM$ and $\TsM$ can be obtained from those in $\TMM$ and $\TsMM$ as $\partial_a:=\Varepsilon_a^\mu\partial_\mu$ and $\eee^a:=\Varepsilon^a_\mu dx^\mu$, where $\Varepsilon_a^\mu:=[{\bm \Varepsilon}^{-1}]_a^\mu$.
Base vectors $\eee^a$ are referred to as the vierbein form, i.e. $\eee\in{V}^1(\TM)\otimes\Omega^1$.
Spin form $\www\in\sss\ooo(1,3)\otimes \Omega^1(\TsM)$ is introduced as a connection-valued one-form in $\TsMM$, which satisfies $\www^t\eta+\eta\www=0$.
The spin form can be represented using a trivial basis as $\www^{ab}=\omega^{~a}_{\mu~c}\hspace{.1em}\eta^{cb}dx^\mu$, where $\omega^{~a}_{\mu~c}:=[\www]^{~a}_{\mu~c}$, and it is antisymmetric as $\www^{ab}=-\www^{ba}$.
We note that the spin form is not a tensor in $\TM$.
A $SO(1,3)$-covariant derivative for $q$-form $\aaa\in\Omega^q$ is defined using the spin form as
\begin{eqnarray*}
d_\www\aaa:=d\aaa+\hspace{.1em}\www\wedge\aaa
+\hspace{.1em}(-1)^{q+1}\aaa\wedge\www.\label{connec2}
\end{eqnarray*}
A direct calculation can show that $d_\www \aaa$ is transformed as a rank-$q$ tensor under $SO(1,3)$ transformations.

Torsion two-form $\TTT^a$ can be defined as the covariant derivative of the vierbein form as follows: 
\begin{eqnarray*}
\TTT^a(\eee,\www)&:=&\td\eee^a=d\eee^a+\www^a_{~\bcdot}\wedge\eee^\bcdot.
\end{eqnarray*}
Hereafter, dummy {\bf Roman} indices are often abbreviated to a small circle (or a star) when the dummy index pair of the Einstein convention is obvious, such as the definition of the torsion form, and this notation is used throughout this study.
By contrast, dummy {\bf Greek} indices are not abbreviated.
When multiple circles appear in an expression, the pairing must be  from left to right for both the upper and the lower indices, e.g. a four-dimensional volume form can be written as 
\begin{eqnarray*}
\vvv&:=&\frac{1}{4!}\epsilon_{\bcdots\bcdots}\eee^\bcdot\wedge\eee^\bcdot\wedge\eee^\bcdot\wedge\eee^\bcdot=\frac{1}{4!}\epsilon_{abcd}\eee^{a}\wedge\eee^{b}\wedge\eee^{c}\wedge\eee^{d}, 
\end{eqnarray*}
where $\epsilon_{abcd}$ is a completely antisymmetric tensor (the Levi-Civita tensor) whose component is $\epsilon_{0123}=1$ in $\TM$.
We note that there are two constant-valued tensors in $\TM$, namely the metric tensor ${\bm \eta}$ and the Levi-Civita tensor ${\bm \epsilon}$, and they have no constant value in $\TMM$.
A two-dimensional surface form is defined as $\SSS_{ab}:=\frac{1}{2}\epsilon_{ab\bcdots}\eee^\bcdot\wedge\eee^\bcdot$, which is a two-dimensional surface perpendicular to both $\eee^a$ and $\eee^b$.
A curvature two-form is defined as 
\begin{eqnarray*}
\RRR^{ab}(d\www,\www)&:=&d\www^{ab}+\www^a_{~\bcdot}\wedge\www^{\bcdot b}. 
\end{eqnarray*}
The spin form $\www^{ab}$ is not a $SO(1,3)$ tensor-valued form, whereas the curvature form $\RRR^{ab}$ is a $SO(1,3)$ tensor-valued two-form.
More precisely, $\RRR^{ab}$ is a rank-$2$ tensor in $\TM$ with respect to the Roman indices and is a two-form in $\TsM$ (and also in $\TsMM$), such as $\RRR\in{V}^2(\TM)\otimes\Omega^2$.
The Bianchi identities can be expressed as
\begin{eqnarray*}
\td\TTT^a=\RRR^{a\bcdot}\wedge\eee^{\bcdot}\hspace{.1em}\eta_{\bcdots},&~~
~&\td\RRR^{ab}=0.
\end{eqnarray*}
A line element is represented as $ds^2:=g_{\mu_1\mu_2}dx^{\mu_1}\otimes dx^{\mu_2}=\eta_\bcdots\eee^\bcdot\otimes\eee^\bcdot$, which is invariant under both  $GL(4)$ and $SO(1,3)$.
Volume form $\vvv$ is also invariant under $GL(4)$ and $SO(1,3)$, and it is expressed as
\begin{eqnarray*}
\vvv&=&\frac{1}{4!}\epsilon_{\bcdott}\Varepsilon^\bcdot_{\mu_1}\Varepsilon^\bcdot_{\mu_2}
\Varepsilon^\bcdot_{\mu_3}\Varepsilon^\bcdot_{\mu_4}\hspace{.1em}
dx^{\mu_1}\wedge dx^{\mu_2}\wedge dx^{\mu_3}\wedge dx^{\mu_4}
=\sqrt{-\det[{\bm g}]}\hspace{.2em}dx^{0}\wedge dx^{1}\wedge dx^{2}\wedge dx^{3},
\end{eqnarray*}
where $\sqrt{-\det[{\bm g}]}=\det[{\bm \Varepsilon}]$.

Einstein--Hilbert gravitational Lagrangian $\LLL$ is expressed using the forms defined above as follows:
\begin{eqnarray}
\LLL_\Lambda
&:=&\frac{1}{\kappa}\left(\RRR\wedge\SSS
-\Lambda_c\vvv
\right),\label{Lag}
\end{eqnarray}
where $\RRR\wedge\SSS=\RRR^\bcdots\wedge\SSS_\bcdots/2$ using a trivial basis, $\kappa=4\pi G$ ($G$ is the Newtonian gravitational constant) is the Einstein gravitational constant in our convention, and $\Lambda_c$ is the cosmological constant.
The light velocity is set to unity as $c=1$, whereas the other natural constants $\kappa$ and $\hbar$ are explicitly expressed.
In this convention, the fundamental parameters are as follows: Planck length $l_p=\sqrt{G\hbar}$, Planck time $t_p=\sqrt{G\hbar}$, and Planck mass $m_p=\sqrt{\hbar/G}$.
We note that $\LLL/\hbar$ has null physical dimension in our definition. 
The Einstein equation and torsionless condition can be expressed as
\begin{eqnarray*}
\frac{1}{2}\epsilon_{a\bcdots\bcdot}
\RRR^\bcdots\wedge\eee^\bcdot
-\Lambda_{\mathrm c}\VVV_a=0,&~&
\TTT^a\wedge\eee^b=0,
\end{eqnarray*}
where $\VVV_a:=\epsilon_{a\bcdot\bcdots}\eee^\bcdot\wedge\eee^\bcdot\wedge\eee^\bcdot/3!$ is a three-dimensional volume form.
They are obtained as an Euler--Lagrange equation by considering the variation with respect to the vierbein and spin forms.
This method is commonly known as the Palatini method\cite{Palatini2008,ferreris}.

In a previous study\cite{doi:10.1063/1.4990708}, the current author introduced a co-Poincar\'{e} symmetry into a space-time manifold. 
While the Einstein--Hilbert gravitational Lagrangian does not have the standard Poincar\'{e} symmetry\cite{0264-9381-29-13-133001}, it has a co-Poincar\'{e} symmetry.
This is a symmetry in which a translation operator is replaced by a co-translation(={\it contraction}+{\it translation}) operator.
Owing to a co-Poincar\'{e} symmetry, the Einstein--Hilbert gravitational Lagrangian is represented using the same shape as the Yang--Mills Lagrangian, and thus, the fundamental forms of general relativity can be identified using an analogy to the Yang--Mills theory. 

A generator of the co-Poincar\'{e} group is expressed as $\left[T_I\right]_{ab}^{cd}=\left(J_{ab},P^{cd}\right)$\footnote{Capital Roman letters indicate indices of a group, and the Einstein convention is also applied to them.}, where $J_{ab}$ and $P^{ab}$ are generators of $SO(1,3)$ and co-translation, respectively.
A generator of co-translation is defined using a trivial basis as $P_{ab}:=P_a\iota_b$, where $\iota_b$ is a contraction operator with respect to trivial basis $\partial_b$.
Co-Poincar\'{e} connection $\AAA_\cP$ and curvature $\FFF_\cP$ are respectively expressed as
\begin{eqnarray*}
\AAA_\cP&:=&T_I\times_\cP\AAA^I_\cP=
\left(J_{ab}, P^{cd}\right)\times_\cP
\left(\vomega^{ab},\SSS_{cd}\right),\\
\FFF_\cP&:=&d\AAA_\cP+\AAA_\cP\wedge\AAA_\cP,
\end{eqnarray*}
where $\times_\cP$ is a product operation with respect to the co-Poincar\'{e} group.
The Lie algebra of a co-Poincar\'{e} group is expressed as
\begin{eqnarray*}
\left[P_{ab},P_{cd}\right]&=&0,\label{PJcr4}\\
\left[J_{ab},P_{cd}\right]&=&-\eta_{ac}P_{bd}+\eta_{bc}P_{ad},\\
\left[J_{ab},J_{cd}\right]&=&
-\eta_{ac}J_{bd}+\eta_{bc}J_{ad}
-\eta_{bd}J_{ac}+\eta_{ad}J_{bc},
\end{eqnarray*}
and it is denoted as $\ggg_\cP$. 

The structure constants $\FF$ of a co-Poincar\'{e} group can be obtained from the above-mentioned Lie algebra through the relation $[T_I,T_J]:=\FF_{~IJ}^{K}\hspace{.1em}T_K$\cite{fre2012gravity}.
Each component of $\FF$ is obtained by direct calculation using a trivial basis as
\begin{eqnarray*}
\left\{
\begin{array}{l}
\FF_{~11}^{1}=\FF_{~11}^{2}=\FF_{~12}^{2}=
\FF_{~21}^{2}=\FF_{~22}^{1}=0,\\
\left[\FF_{~12}^{1}\right]_{ab;cd}^{ef}=
-\left[\FF_{~21}^{1}\right]_{cd;ab}^{ef}=
\eta_{ac}\delta^e_b\delta^f_d-\eta_{bc}\delta^e_a\delta^f_d,\\
\left[\FF_{~22}^{2}\right]_{ab;cd}^{ef}=
-\eta_{ac}\delta_b^e\delta_d^f+\eta_{bc}\delta_a^e\delta_d^f
-\eta_{bd}\delta_a^e\delta_c^f+\eta_{ad}\delta_b^e\delta_c^f.
\end{array}
\right.\label{LiecoP}
\end{eqnarray*}
From the above-mentioned expressions of the structure constant, co-Poincar\'{e} connection and curvature forms are obtained as
\begin{eqnarray}
\AAA_\cP&=&
J_\bcdots\vomega^\bcdots+ P^\bcdots\SSS_\bcdots\hspace{1.3em}
\in\Omega^1\otimes\ggg_\cP,\label{cPB1}\\
\FFF_\cP&=&J_\bcdots\RRR^\bcdots+P^\bcdots\td\SSS_\bcdots
\in\Omega^2\otimes\ggg_\cP.\label{cPB2}
\end{eqnarray}
The Einstein--Hilbert Lagrangian form has topological invariance under the general coordinate transformation and the co-Poincar\'{e} transformation:
\begin{remark}\label{rem2.1}
The Einstein--Hilbert Lagrangian without the cosmological constant is represented using the curvature of the co-Poincar\'{e} bundle $(${\bf Theorem 4.2} in Ref.\cite{doi:10.1063/1.4990708}$)$ as
\begin{eqnarray}
\LLL&:=&\frac{1}{\kappa}\RRR\wedge\SSS=
-\frac{1}{\kappa}
\Tr\left[\FFF_\cP\wedge\FFF_\cP\right].\label{gravact}
\end{eqnarray}
It has a topological invariant as the second Chern class $c_2(\FFF_\cP)$: 
\begin{eqnarray}
\LLL=
\frac{8\pi^2}{\kappa}c_2(\FFF_\cP)\in \R\otimes H^4(\M,\Z).
\end{eqnarray}
\end{remark}
\begin{proof}
The right-hand side of (\ref{gravact}) is expressed using the structure constants as
\begin{eqnarray*}
\Tr\left[\FFF_\cP\wedge\FFF_\cP\right]&:=&\frac{1}{2}
\FF^K_{~IJ}\hspace{.1em}\FFF_\cP^I\wedge\FFF_\cP^J\hspace{.1em}T_K. 
\end{eqnarray*}
From representation (\ref{cPB2}) and the Lie algebra of the co-Poincar\'{e} symmetry, direct calculations using a trivial basis yield
\begin{eqnarray*}
\Tr\left[\FFF_\cP\wedge\FFF_\cP\right]&=&
\Tr\left[\RRR^{ac}\wedge
d_\www\left(P_{a}^{\hspace{.3em}b}\hspace{.1em}\SSS_{cb}\right)\right]
=
d_\www\hspace{-.2em}\left(\Tr\left[P_{a}^{\hspace{.3em}b}\right]
\RRR^{ac}\wedge\SSS_{cb}\right)=
-d_\www\hspace{-.2em}
\left(P_{\bcdot}^{\hspace{.3em}\bcdot}
\hspace{.1em}\RRR\wedge\SSS\right),
\end{eqnarray*}
where Bianchi identity $d_\www\RRR=0$ and co-translation invariance $P\hspace{.1em}(\RRR)=0$ are used.
For the last equality, we note that $\Tr\left[\RRR^{ac}\wedge\SSS_{cb}\right]=-\delta^a_b(\RRR\wedge\SSS)$.
The Einstein--Hilbert gravitational Lagrangian is co-translation-invariant up to total derivative\footnote{{\bf Remark 3.2} in ref\cite{doi:10.1063/1.4990708}}; thus, $P$ is removed from the last expression and it is embedded in a five-dimensional line bundle of $\MM_5:=\MM\otimes\R$, whose boundary is given as $\partial\MM_5=\Sigma$.
Consequently,  
\begin{eqnarray*}
-\int_{\Sigma}\Tr\left[\FFF_\cP\wedge\FFF_\cP\right]&=&
\int_{\MM_5}d_\www\left(\RRR\wedge\SSS\right)=
\int_{\MM_5}d\left(\RRR\wedge\SSS\right)=
\int_{\partial\hspace{-.1em}\MM_5=\Sigma}\RRR\wedge\SSS.
\end{eqnarray*}
A surface term is eliminated owing to a boundary condition.
We note that 
\begin{eqnarray*}
d_\www\left(\RRR\wedge\SSS\right)=
d\left(\RRR\wedge\SSS\right)
+\www\wedge\RRR\wedge\SSS
-\RRR\wedge\SSS\wedge\www=d\left(\RRR\wedge\SSS\right),
\end{eqnarray*}
owing to the definition of a covariant derivative.

The second Chern class with respect to the co-Poincar\'{e} curvature is obtained as  
\begin{eqnarray*}
c_2(\FFF_\cP)&:=&\frac{1}{8\pi^2}\left(\Tr\left[\FFF_\cP\right]^2
-\Tr\left[\FFF_\cP\wedge\FFF_\cP\right]\right),
\end{eqnarray*}
and it has a homology class as $c_2(\FFF_\cP)\in H^4(\M,\R)\subset\mathrm{Im}\hspace{.1em}H^4(\M,\Z)$ owing to the Chern--Weil theory.
Because the co-Poincar\'{e} curvature $\FFF_\cP$ is traceless, {\bf Remark \ref{rem2.1}} holds.
\end{proof}
\noindent
This result suggests that appropriate fundamental forms (phase space) of the symplectic geometry for general relativity can be identified as $(\www,\SSS)$.
More precisely, we propose $(4!)$-pairs of phase space variables: 
\begin{eqnarray*}
(\www,\SSS)&\sim&
(\left[\www\right]^{a_1a_2}_{~~\mu_1},
\left[\SSS\right]_{b_1b_2;\mu_2\mu_3})
\sim
(\left[\www\right]^{a_1a_2},
\left[\SSS\right]_{b_1b_2})_{\mu_1\mu_2\mu_3},
\end{eqnarray*}
where each pair of variables has $(6\hspace{-.1em}+\hspace{-.1em}6)$-degrees of freedom (DoF).
We noted that not all of the 24 pairs are independent of each other.
This choice of the canonical pair of general relativity is equivalent to that proposed by Kanatchikov\cite{Kanatchikov_2013} on the basis of the de\hspace{.1em}Donder--Weyl Hamiltonian theory.

%
%
\subsection{Hamiltonian formalism and Hamilton--Jacobi equation}\label{HamiltonJacobi}
Our choice of fundamental forms corresponds to real polarization in which half of the principal (co-Poincar\'{e}) connection is chosen for a configuration variable.
Further details are provided later in this report.
When spin form $\www$ is identified as the first fundamental variable (the general configuration variable), the second canonical variable $\MMM$ (canonical momentum) is defined according to the standard definition as
\begin{eqnarray*}
\MMM_{ab}:=\frac{1}{\hbar}\frac{\delta{\LLL}}{\delta\left(d\vomega^{ab}\right)}=\frac{1}{\kappa\hbar}\SSS_{ab}.
\end{eqnarray*}
The Planck constant $\hbar$ is introduced to adjust the physical dimension.
The surface form has physical dimension $(length)^2$; thus, $\SSS/(\kappa\hbar)$ has null physical dimension, as in the case of $\www$.
Even though $\hbar$ appears in the equations, the classical theory is still treated in this section.
A canonical momentum is identified as a surface form; thus, the fundamental forms are $(\www,\SSS)$, which are consistent with the principal bundle of the co-Poincar\'{e} symmetry.

Through the Legendre transformation, a classical Hamiltonian form can be obtained from the Lagrangian form as
\begin{eqnarray}
\HHH&=&
\frac{1}{2}\MMM_\bcdots\wedge d\www^\bcdots-\frac{1}{\hbar}{\LLL}=
-\frac{1}{2\kappa\hbar}
\vomega^{\star}_{~\bcdot}\wedge\vomega^{\bcdot\star}\wedge\SSS_{\stars}.\label{CHG}
\end{eqnarray}
As mentioned at the end of previous section, fundamental forms, $\www$ and $\SSS$, are not independent each other, and thus, the Hamiltonian system has constraints. 
A standard method to treat a constrained system is to add constrained conditions with the Lagrangian multipliers to the Hamiltonian, and then, fixing the Lagrangian multipliers using a canonical equations.
The Hamiltonian (\ref{CHG}) is given as that after eliminating the Lagrange multipliers according to a standard method (see section 2 of Ref.\cite{Kanatchikov_2013}).

The Hamiltonian form has null physical dimension in our convention.
We note that the Hamiltonian form is not the local ${SO}(1,3)$ invariant because it consists of the spin form itself, which is not a Lorentz tensor.
This is natural because the gravitational field can be eliminated at any point in $\MM$ owing to the Einstein equivalence principle; thus, the gravitational energy depends on the local observer.
For example, the energy of a gravitational wave is well defined only on an asymptotically flat space-time manifold (see, for example, section 11.2 of Ref.\cite{wald2010general}).

The first canonical equation can be obtained as
\begin{eqnarray*}
\frac{\delta\HHH_G}{\delta\MMM_{ab}}&=&
-\epsilon_{\bcdott}\left(\epsilon_{ab\bcdot\star}\eee^\star\right)^{-1}\wedge\eee^\bcdot\wedge\www^\bcdot_{~\star}\wedge\www^{\star\bcdot}=
d\vomega^{ab},~~
\end{eqnarray*}
where $\delta\bullet/\delta\SSS=(\delta\bullet/\delta\eee)(\delta\SSS/\delta\eee)^{-1}$ is used.
Thus, we can obtain the first equation of motion as
\begin{eqnarray}
-\epsilon_{a\bcdots\bcdot}\eee^\bcdot\wedge\www^\bcdot_{~\star}
\wedge\www^{\star\bcdot}&=&
\epsilon_{a\bcdots\bcdot}\eee^\bcdot\wedge d\vomega^{\bcdots}.\label{canonicaleq1}
\end{eqnarray}
The second canonical equation can be obtained as
\begin{eqnarray}
\frac{\delta\HHH_G}{\delta\vomega^{ab}}&=&-d\MMM_{ab},\label{canonicaleq2}
\end{eqnarray}
Here, (\ref{canonicaleq1}) provides the equation of motion
$
\epsilon_{a\bcdots\bcdot}\RRR^\bcdots\wedge\eee^\bcdot=0,
$
which simply leads to the Einstein equation without matter fields nor the cosmological constant, and (\ref{canonicaleq2}) provides
\begin{eqnarray*}
\td\SSS_{a\bcdot}\wedge\eee^\bcdot&=&2\TTT^\bcdot\wedge\SSS_{a\bcdot}=0,
\end{eqnarray*}
Which leads to the torsionless condition, as expected. 

The Poisson bracket is introduced in the covariant formalism as
\begin{eqnarray}
\left\{\aaa,\bbb\right\}_\mathrm{PB}&=&
\frac{\delta\aaa}{\delta\vomega^\bcdots}\wedge\frac{\delta\bbb}{\delta\SSS_\bcdots}-
\frac{\delta\bbb}{\delta\vomega^\bcdots}\wedge\frac{\delta\aaa}{\delta\SSS_\bcdots},\label{PB}
\end{eqnarray}
where $\aaa\in\Omega^p$ and $\bbb\in\Omega^q$.
The Poisson bracket can be recognized as a map:
\begin{eqnarray*}
\{\bullet,\bullet\}_\mathrm{PB}:\Omega^p\otimes\Omega^q\rightarrow\Omega^{p+q-3}:(\aaa,\bbb)\mapsto(\ref{PB}).
\end{eqnarray*}
The Poisson brackets for the fundamental forms are obtained as
\begin{eqnarray*}
\left\{\vomega^{a_1a_2},\vomega^{a_3a_4}\right\}_\mathrm{PB}&=&
\left\{\SSS_{b_1b_2},\SSS_{b_3b_4}\right\}_\mathrm{PB}~=~0,\\
\left\{\vomega^{a_1a_2},\SSS_{b_1b_2}\right\}_\mathrm{PB}&=&
\delta^{[a_1}_{b_1}\delta^{a_2]}_{b_2},\label{pb2}
\end{eqnarray*}
where $\delta^{[a_1}_{b_1}\delta^{a_2]}_{b_2}=\delta^{a_1}_{b_1}\delta^{a_2}_{b_2}-\delta^{a_2}_{b_1}\delta^{a_1}_{b_2}$.
The Hamiltonian form can be understood as a generator of the total derivative of a given form because the Poisson brackets of the fundamental forms and the Hamiltonian form are expressed as 
\begin{eqnarray*}
\epsilon_{a\bcdots\bcdot}\left\{\vomega^\bcdots,\HHH\right\}_\mathrm{PB}\wedge\eee^\bcdot&=&
-\epsilon_{a\bcdots\bcdot}\vomega^\bcdot_{~\star}\wedge\vomega^{\star\bcdot}\wedge\eee^\bcdot
=\epsilon_{a\bcdots\bcdot}d\vomega^\bcdots\wedge\eee^\bcdot,\\
\left\{\SSS_{ab},\HHH\right\}_\mathrm{PB}&=&
-\left(-\eta_{b\bcdot}\vomega^\bcdots\wedge\SSS_{\bcdot a}\right)=d\SSS_{ab},
\end{eqnarray*}
where the canonical equations of motion are used.
These results are consistent with those presented in Ref.\cite{Kanatchikov_2013}.

Lagrangian form $\LLL$ is a four-form object defined on four-dimensional manifold $\M$; thus, it is a closed form, i.e. $d\LLL=0$.
Three-form $\GGG$, which satisfies $\LLL=d\GGG$, exists at least locally in simply connected submanifold $\Sigma\subset\M$ owing to the Poincar\'{e} lemma.
Three-form $\GGG$ is referred to as the generating form hereafter.
The existence of the generating form in $\Sigma$ is assumed here.

A general coordinate transformation is a smooth automorphism on $\Sigma$ as $\Gamma:\Sigma\rightarrow\Sigma$, where map $\Gamma$ is a canonical transformation because the Hamiltonian equations are invariant under $\Gamma\in{GL}(1,3)$. 
A pull-back of generating form $\Gamma^\sharp\GGG$ provides the same action integral as $\GGG$ does, i.e. 
\begin{eqnarray*}
\int_{\Sigma}\LLL&=&\int_{\Sigma} d\GGG=\int_{\Sigma} d\left(\Gamma^\sharp\GGG\right).
\end{eqnarray*}
For the generating form, pull-back $\Gamma^\sharp\GGG$ is equated to the original generating form and is denoted using the same symbol as $\Gamma^\sharp\GGG=\GGG$.
The generating form is a functional of $\www$ and $d\www$ as well as Lagrangian form $\LLL(\www,d\www)$.

Under these observations, the Hamilton--Jacobi equation of general relativity can be obtained as follows:
A canonical transformation of the Hamiltonian (\ref{CHG}) is provided as
\begin{eqnarray*}
\widetilde{\HHH}&=&\HHH(\twww,\tSSS)=
\frac{1}{2\kappa\hbar}\tSSS_\bcdots\wedge d\twww^\bcdots-\frac{1}{\hbar}\widetilde\LLL,
\end{eqnarray*}
where $\widetilde\bullet:=\Gamma^\sharp\bullet$.
The spin form is not a Lorentz tensor; thus, the Hamiltonian form and the first term on the right-hand side of (\ref{CHG}) depend on the local coordinate frame. 
However, the Lagrangian form is invariant under canonical transformation $\widetilde\LLL=\LLL$; thus, the relationship
\begin{eqnarray}
\frac{1}{\hbar}\LLL&=&\frac{1}{\hbar}d\GGG
=\frac{1}{2\kappa\hbar}\SSS_\bcdots\wedge d\www^\bcdots-\HHH(\www,\SSS)
=\frac{1}{2\kappa\hbar}\tSSS_\bcdots\wedge d\twww^\bcdots-\HHH(\twww,\tSSS),\label{HJ1}
\end{eqnarray}
is obtained.
Therefore, we have the following formal expression:
\begin{eqnarray}
\frac{\delta\GGG}{\delta\www^{ab}}=\frac{1}{\kappa}\SSS_{ab},&~~&
\frac{\delta\GGG}{\delta\twww^{ab}}=\frac{1}{\kappa}\tSSS_{ab}.\label{HJ2}
\end{eqnarray}
From (\ref{HJ1}) and (\ref{HJ2}), the Hamilton--Jacobi equation for general relativity is given as
\begin{eqnarray}
\frac{1}{\hbar}d\GGG&=&\frac{1}{2\hbar}\frac{\delta\GGG}{\delta\www^\bcdots}\wedge d\www^\bcdots-
\HHH\left(\www,\kappa\frac{\delta\GGG}{\delta\www}\right)=
\frac{1}{\hbar}d\GGG-
\HHH\left(\www,\kappa\frac{\delta\GGG}{\delta\www}\right),
\nonumber\\
&\Longrightarrow& \HHH\left(\www,\kappa\frac{\delta\GGG}{\delta\www}\right)=0,
\end{eqnarray}
where $\delta\GGG/\delta(d\www^\bcdots)\wedge d(d\www^\bcdots)=0$ is used.

The Einstein equation without any matter or gauge fields is considered, and its classical solutions are referred to as the ``vacuum solution'', denoted by the suffix ``$vac$''.
In this case, the Lagrangian form itself is zero, i.e. $\LLL(\www_{\hspace{-.1em}vac}, d\www_{\hspace{-.1em}vac})=0$ because of the equation of motion.
From (\ref{HJ1}), the Hamiltonian form can be expressed as
\begin{eqnarray}
\HHH\left(
\www_{\hspace{-.1em}vac},\SSS^{\hspace{-.1em}vac}
\right)&=&\frac{1}{2\kappa\hbar}\SSS^{\hspace{-.1em}vac}_\bcdots\wedge
d\www^\bcdots_{\hspace{-.1em}vac}.\label{HSW}
\end{eqnarray}
We note that the four-form item on the right-hand side of (\ref{HSW}) is not a functional integration with respect to the spin form but is a Riemannian integration.
Two-form $d\www^{ab}_{\hspace{-.1em}vac}$ does not mean a functional integration measure but simply denotes an external derivative of the spin form.
This four-form item can be represented using a trivial basis in $\TsMM$ as
\begin{eqnarray*}
\SSS^{{\hspace{-.1em}vac}}_\bcdots\wedge d\www^\bcdots_{{\hspace{-.1em}vac}}
&=&\epsilon_{\bcdots\bcdots}\hspace{.2em}
\Varepsilon^\bcdot_{\mu_1}
\Varepsilon^\bcdot_{\mu_2}
\left(\partial_{\mu_3}\omega^{~~\bcdots}_{\mu_4}\right)
dx^{\mu_1}\wedge dx^{\mu_2}\wedge dx^{\mu_3}\wedge dx^{\mu_4},
\end{eqnarray*}
where suffix ``${\hspace{-.1em}vac}$'' is omitted from vierbein $\Varepsilon^a_\mu$ and spin connection $\omega^{\hspace{.4em}ab}_{\mu}$ for simplicity.
The Hamiltonian form is four-form $\HHH\in\Omega^4$, and four-forms in a four-dimensional manifold have only one base vector, i.e. the volume form.
Therefore, that Hamiltonian form is proportional to the volume form $\vvv$, and it can be written as $\HHH(\www_{{\hspace{-.1em}vac}},\SSS^{{\hspace{-.1em}vac}})=H_{{\hspace{-.1em}vac}}(x)\vvv/(\kappa\hbar)$, where $H_{{\hspace{-.1em}vac}}(x)$ is a dimensionless energy density function.
The total energy in compact orientable four-dimensional manifold $\Sigma$ can be obtained as
\begin{eqnarray}
E_{{\hspace{-.1em}vac}}:=\frac{1}{\kappa\hbar}\int_\Sigma H_{{\hspace{-.1em}vac}}(x)\vvv.\label{scH}
\end{eqnarray}
Energy density function $H_{{\hspace{-.1em}vac}}(x)$ is not a Lorentz scalar; thus, total energy $E_{{\hspace{-.1em}vac}}$ depends on a local observer.

\section{Symplectic structure of general relativity}
The symplectic geometry originated from studies of the Hamiltonian formalism of classical mechanics. 
The Hamiltonian formalism of classical general relativity was introduced in the previous section.
A symplectic manifold is defined as follows:
\begin{definition}{\bf (symplectic manifold)}\\
In $2n$-dimensional real manifold $M$, $(l\hspace{-.1em}+\hspace{-.1em}2)$-form $\WWW$ is considered, where $l\in\N$ is the total rank of phase-space variables as differential forms in $\TM$.
When $\WWW$ is non-degenerate and satisfies $d\WWW=0$, pair $(M,\WWW)$ is called a symplectic manifold, and $\WWW$ is called a symplectic form.
If $(l\hspace{-.1em}+\hspace{-.1em}1)$-form $\sss$ exists such that $\WWW=d\sss$, $\sss$ is called a Liouville form. 
\end{definition}
\noindent
Classical general relativity has a symplectic structure, which is stated as the following theorem:
\begin{theorem}\label{sypform}
Classical general relativity has a symplectic structure with respect to the Liouville form $\sssgr:=\frac{1}{2\kappa\hbar}\SSS_\bcdots\wedge d\www^\bcdots$.
\end{theorem}
\noindent
The first objective of this study is to prove {\bf Theorem \ref{sypform}}.

%
%
\subsection{Proof of {\bf Theorem \ref{sypform}}}\label{2.1}
We consider form $\WWWgr:=d\sssgr=\frac{1}{2}d\SSS_\bcdots\wedge d\www^\bcdots$, where fundamental physical constants $\kappa$ and $\hbar$ are omitted in this section.
To prove the theorem, we must show that (1) manifold $M=\Mgr$ exists, (2) form $\WWWgr$ is non-degenerate, and (3) form $\WWWgr$ is closed.
Among these three statements, the third one is trivial owing to its definition.
The proofs of statements (1) and (2) are provided below.

\subsubsection{Existence of $\Mgr$}\label{EofMgr}
A set of square-integrable real functions is introduced as a section in $\MM$, which is denoted as $L^2(\MM)$.
This space of functions is considered as a Hilbert space, and it is denoted as $\HH$.
Spin forms, obtained as solutions to the Einstein equation, are assumed to be square-integrable; thus, the space of the coefficient functions for the spin forms, denoted as ``$\varpi$'', is a subset of the Hilbert space: $\varpi\subset\HH_\omega\subset L^2$.
Here, a new operator is introduced as a mapping operator that transforms a rank-$p$ tensor into a rank-$(n\hspace{-.2em}-\hspace{-.2em}p)$ tensor, which is defined such that
\begin{eqnarray}
\overline{\aaa^{a_1\cdots a_p}}&=&\overline{\aaa}_{a_1\cdots a_{n-p}}:=~~
\frac{1}{p!}\epsilon_{a_1\cdots a_{n-p} b_1\cdots b_p}
\aaa^{b_1\cdots b_p}\label{barop1},\\
\overline{\aaa_{a_1\cdots a_p}}&=&\overline{\aaa}^{a_1\cdots a_{n-p}}:=-\frac{1}{p!}
\epsilon^{b_1\cdots b_pa_1\cdots a_{n-p} }
\aaa_{b_1\cdots b_p}\label{barop1},
\end{eqnarray}
where $\aaa\in\Omega^q\oplus V^p(\TM)$ is an arbitrary $q$-form of a rank-$p$ tensor and $\epsilon^{a_1\cdots a_n}=[{\bm \epsilon}]^{a_1\cdots a_n}$ is a completely antisymmetric tensor (the Levi-civita tensor) in an $n$-dimensional space, which has been already identified for a four-dimensional space in Section 1.
We note that the raising and lowering Roman indices (tensorial indicies defined in $\M$) is carried out using a metric tensor $\bm\eta$ in $\M$.
Here, this operator is referred to as the bar-dual operator.
It follows from the definition of the relationship $\overline{\overline{\aaa}}=\aaa$ when $\aaa$ is antisymmetric with respect to all the indices. 
For instance, surface- and volume-forms introduced in Section 1 can be expressed using the bar-dual operator as 
\begin{eqnarray*}
\SSS_{ab}&=&\overline{\eee^\bullet\wedge\eee^\bullet}=
\frac{1}{2!}\epsilon_{ab\bcdots}\eee^\bcdot\wedge\eee^\bcdot,\\
\VVV_a&=&\overline{\eee^\bullet\wedge\eee^\bullet\wedge\eee^\bullet}=
\frac{1}{3!}\epsilon_{a\bcdot\bcdots}\eee^\bcdot\wedge\eee^\bcdot\wedge\eee^\bcdot,\\
\vvv&=&\overline{\eee^\bullet\wedge\eee^\bullet\wedge\eee^\bullet\wedge\eee^\bullet}=
\frac{1}{4!}\epsilon_{\bcdots\bcdots}\eee^\bcdot\wedge\eee^\bcdot\wedge\eee^\bcdot\wedge\eee^\bcdot,
\end{eqnarray*}
using (\ref{barop1}).
We note that the bar-dual operator is different from the Hodge-$*$ operator.
In contrast to the Hodge-$*$ operator, which transforms a $p$-form into an $(n\hspace{-.2em}-\hspace{-.2em}p)$-form, the bar-dual operator transforms a rank-$p$ tensor to a rank-$(n\hspace{-.2em}-\hspace{-.2em}p)$ tensor while keeping the rank of a form; $\bar{\bullet}:\Omega^q\oplus V^p(\TM)\rightarrow\Omega^q\oplus V^{n-p}(\TM)$.

The bar-dual space of $\varpi$ is denoted as $\overline{\varpi}$ hereafter.
The Gel'fand triple with respect to the bar-dual operator becomes  $\varpi\subseteq\HH_\omega\subseteq\overline{\varpi}$.
For spin form $\www^{ab}$ and its bar-dual form $\overline\www^{ab}$, their coefficient functions are expressed as $\omega_{\mu}^{\hspace{.4em}ab}\in\varpi$ and $\frac{1}{2}\epsilon_{ab\bcdots}\hspace{.2em}\omega_{\mu}^{\hspace{.4em}\bcdots}\in\overline\varpi$, respectively.
The element of bar-dual space $\overline{\varpi}$ is a linear combination of spin forms $\omega^{ab}_{~~\mu}\in\varpi$; thus,  functional spaces $\varpi$ and $\overline{\varpi}$ are isomorphic: $\varpi\cong \overline{\varpi}$.
Therefore, the Hilbert space can be considered as $\HH_\omega=\varpi$ because the Gel'fand triple is now  $\varpi\subseteq\HH_\omega\subseteq \varpi$.
Standard bilinear form $\langle \overline{\www}|\www\rangle$ is defined as
\begin{eqnarray*}
\langle \overline\www|\www\rangle&:=&\frac{1}{2}\overline\www_\bcdots\wedge\www^\bcdots=
\frac{1}{8}\epsilon_{\bcdott}\hspace{.2em}
\omega^{~~\bcdots}_{\mu_1}\hspace{.2em}\omega^{~~\bcdots}_{\mu_2}\hspace{.2em}
dx^{\mu_1}\wedge dx^{\mu_2}.
\end{eqnarray*}
The norm of spin form $\|\www\|$ can be defined using the bilinear form as
\begin{eqnarray*}
\|\www\|^2:=\int_{\Sigma_2}\langle\overline\www|\www\rangle\in\R,
\end{eqnarray*}
where ${\Sigma_2}$ is an appropriate two-dimensional sub-manifold ${\Sigma_2}\subset\M$.
Further, $d\www^{ab}=d\omega^{~ab}_\mu dx^\mu$ is a two-form defined in $\TsM$, and its coefficient function $d\omega^{~ab}_\mu=(\partial_\bcdot\omega^{~ab}_\mu)\eee^\bcdot\in\Omega^1(T^*\hspace{-.1em}\varpi)$ is a one-form defined in $T^*\hspace{-.1em}\varpi$.

The surface form and its bar-dual form can be respectively expressed using a trivial basis in $\TsMM$ as
\begin{eqnarray*}
\SSS_{ab}&:=&S_{ab;\mu_1\mu_2}\hspace{.2em}dx^{\mu_1}\wedge dx^{\mu_2},\\
\overline{\SSS_{ab}}&=&\eee^a\wedge\eee^b:=\overline{S}^{\hspace{.2em}ab}_{\hspace{.4em}\mu_1\mu_2}\hspace{.2em}dx^{\mu_1}\wedge dx^{\mu_2},
\end{eqnarray*}
where
\begin{eqnarray*}
S_{ab;\mu_1\mu_2}=\frac{1}{2}\epsilon_{ab\bcdots}\hspace{.2em}
\Varepsilon^\bcdot_{\mu_1}\hspace{.2em}\Varepsilon^\bcdot_{\mu_2}~~&\mathrm{and}&~~
\overline{S}^{\hspace{.2em}ab}_{\hspace{.4em}\mu_1\mu_2}=
\Varepsilon^a_{\mu_1}\hspace{.2em}\Varepsilon^b_{\mu_2}.
\end{eqnarray*}
A space in component functions $\sigma:=\{S\}=\{\overline{S}\}=:\overline\sigma$ is square-integrable and forms Gel'fand triple $\sigma\subseteq\HH_\SSS\subseteq\overline{\sigma}=\sigma$; thus, $\HH_\SSS=\sigma$ is maintained.
The standard bilinear form is introduced to be the same as that of the spin form:
\begin{eqnarray*}
\langle \overline\SSS|\SSS\rangle&:=&\frac{1}{2}\overline\SSS^\bcdots\otimes\SSS_\bcdots=
\frac{1}{4}\epsilon_{\bcdott}\hspace{.2em}
\Varepsilon^{\bcdot}_{\mu_1}\hspace{.2em}
\Varepsilon^{\bcdot}_{\mu_2}\hspace{.2em}
\Varepsilon^{\bcdot}_{\mu_3}\hspace{.2em}
\Varepsilon^{\bcdot}_{\mu_4}\hspace{.2em}
dx^{\mu_1}\wedge dx^{\mu_2}\wedge dx^{\mu_3}\wedge dx^{\mu_4}
=(3!)\hspace{.1em}\vvv.
\end{eqnarray*}
The norm of surface form $\|\SSS\|$ can be defined as
\begin{eqnarray*}
\|\SSS\|^2:=\int_{\Sigma_4}\langle\overline\SSS|\SSS\rangle=3!\int_{\Sigma_4}\vvv
\in\R,
\end{eqnarray*}
where ${\Sigma_4}$ is an appropriate four-dimensional submanifold ${\Sigma_4}\subset\M$.

Vierbein $\Varepsilon^a_\mu$ is a $(4\hspace{-.2em}\times\hspace{-.2em}4)$-matrix for maintaining $SO(1,3)$-symmetry; thus, it has a total of $4\times4-6=10$ DoF.
The tensor coefficient of spin form $\omega_{\mu}^{\hspace{.4em}ab}$ is antisymmetric with respect to the Roman indices, and it has a total of $4\times(4\times3)/2=24$ DoF.
In addition, the torsionless equations can be represented using a trivial basis as
\begin{eqnarray*}
\left(\partial_\mu\Varepsilon^a_\nu+\omega^{~a}_{\mu~\bcdot}\hspace{.2em}\Varepsilon^\bcdot_\nu\right)
dx^\mu\wedge dx^\nu&=&0,
\end{eqnarray*}
which include 24 independent equations.
Therefore, when all 10 independent components of the vierbein are given, the spin connection can be uniquely fixed.
Thus, the spin form has a total of 10 DoF, which is the same as that of the vierbein.
The surface form is also uniquely fixed from the vierbein form; thus, it has a total of 10 DoF.
The pairs of fundamental forms $(\www,\SSS)$ have a total $20$-dimensional functional spaces, and not all of them are independent.
Given that the vierbein form $\eee^a$ creates an orthogonal base in $\TsM$, the spin and surface forms can be expanded using the vierbein form (the surface form is described by its definition).
Therefore, once a solution of the Einstein equation is obtained by means of the vierbein form, both the spin and the surface forms can be expressed using this vierbein; thus, $\varpi\otimes\sigma$ is not empty.

Consequently, $20$-dimensional base manifold $\Mgr$ of symplectic manifold $(\Mgr,\WWWgr)$ is constructed as a Hilbert manifold as follows: $\Mgr:=\HH_\omega\otimes\HH_\SSS=\varpi\otimes\sigma$.
We note that any Hilbert space $\HH$ can be regarded as a manifold, i.e. a Hilbert manifold, with respect to the identity function of $\HH$.
This Hilbert space is called a prequantum Hilbert space in the context of geometrical quantization.

\subsubsection{Non-degeneracy of $\WWWgr$}\label{nondege}
In four-dimensional manifold $\M_4$, $\WWWgr$ itself is identically zero because $\WWWgr$ is a five-form item; thus, it exhibits an empty theory (subscript ``$4$'' is introduced to the four-dimensional manifold, i.e. $\M_4$, to distinguish it from the five-dimensional one in this section).
Non-empty theory is constructed by embedding $\WWWgr$ in a line-bundle: $\M_5:=\M_4\otimes\R$.

The symplectic form embedded in five-dimensional manifold $\Sigma_5\subset\M_5\in\left(\M_4\otimes\R\right)$ is considered with the boundary as  $\partial\Sigma_5:=\Sigma_4\subset\Sigma_5$.
Manifold $\Sigma_5$ is assumed to have no boundary other than $\Sigma_4$.
Inclusion map $\varphi_\rho:\M_4\rightarrow\M_5$ maps non-singular $(4\hspace{-.2em}\times\hspace{-.2em}4)$-matrix (rank-$2$ tensor) ${\bm T}_4$ to non-singular $(5\hspace{-.2em}\times\hspace{-.2em}5)$-matrix (rank-$2$ tensor) ${\bm T}_5$ as follows:
\begin{eqnarray*}
{\bm T}_5&:=&\varphi_\rho\left({\bm T}_4\right)=
 \left[
    \begin{array}{cc}
      {\bm T}_4 & {\bm 0}\\
       {\bm 0}^t & \rho
    \end{array}
  \right],
\end{eqnarray*}
where $\rho\in\R$ is an arbitrary constant and ${\bm 0}:=(0,0,0,0)^t$ is a zero vector.
Point $\xi$ on $\Sigma_4$ is transformed into $\varphi_\rho[(\xi^0,\xi^1,\xi^2,\xi^3)]=(\xi^0,\xi^1,\xi^2,\xi^3,\rho)$ on each chart.
Manifold $\M_5$ is covered by $\varphi_\rho(\M_4)$ as  $\M_5=\bigcap_{\rho\in\R}\varphi_\rho(\M_4)$, where $\varphi_\rho\cup\varphi_{\rho'}=O\hspace{-0.7em}/\hspace{0.5em}$ if $\rho\neq\rho'$; thus, map $\varphi_\rho$ induces a foliation of $\M_5$, whose leaf is $\M_4$.
The derivative of $\varphi_\rho$ acts on ${\bm T}_4$ as
\begin{eqnarray*}
D\varphi_\rho[{\bm T}_4]&=&\left(\varphi_\rho[{\bm T}_4+{\bm \tau}]-\varphi_\rho[{\bm T}_4]\right){\bm \tau}^{-1}=
 \left[
    \begin{array}{cc}
      {\bm I}_4 & {\bm 0}\\
      {\bm 0}^t & 0 
    \end{array}
  \right],
\end{eqnarray*}
where ${\bm \tau}$ is any non-singular $(4\hspace{-.1em}\times\hspace{-.1em}4)$-matrix and ${\bm \tau}^{-1}$ is its inverse, and ${\bm I}_4$ is a $(4\hspace{-.1em}\times\hspace{-.1em}4)$-identity matrix.
Given that map $D\varphi_\rho$ maintains the rank of a matrix, map $\varphi_\rho$ is immersed; thus, $D\varphi_\rho$ is a map of the tangent bundle as $D\varphi_\rho:\TM_4\rightarrow\TM_5$.

The integral of form $\WWWgr$ in $\Sigma_5$ is given as
\begin{eqnarray}
\int_{\Sigma_5}\WWWgr&=&\frac{1}{2}\int_{\Sigma_5}d\SSS_\bcdots\wedge d\www^\bcdots
=\frac{1}{2}
\int_{\Sigma_5}d\left(\SSS_\bcdots\wedge d\www^\bcdots\right)
=\frac{1}{2}
\int_{\partial\Sigma_5=\Sigma_4}\varphi_\rho^\sharp\left(\SSS_\bcdots\wedge d\www^\bcdots\right),\label{S5}
\end{eqnarray}
where Storks' theorem is used, and map $\varphi$ maintains the relationship
\begin{eqnarray*}
(\ref{S5})&=&
\frac{1}{2}\int_{\Sigma_4}\frac{1}{2}\epsilon_{\bcdots\bcdots}\eee^\bcdot\wedge\eee^\bcdot
\wedge d\www^\bcdots=\Sgr.
\end{eqnarray*}
Therefore, the non-degeneracy of $\WWWgr$ in $\Sigma_5$ is equivalent to that of $\sssgr=\SSS_\bcdots\wedge d\www^\bcdots/2$ in $\Sigma_4$.
Liouville form $\sssgr$ can be written using a trivial basis as 
\begin{eqnarray*}
\sssgr&=&\frac{1}{4}\epsilon_{a_1a_2a_3a_4}\left(\partial_{b_1}\omega^{~~a_1a_2}_{b_2}\right)\hspace{.2em}
\eee^{a_3}\wedge\eee^{a_4}\wedge\eee^{b_1}\wedge\eee^{b_2}=
\frac{1}{2}\left(\partial_{b_1}\omega^{~~a_1a_2}_{b_2}\right)\delta^{[b_1}_{a_1}\delta^{b_2]}_{a_2}\hspace{.2em}\vvv,
\end{eqnarray*}
where $\omega_a^{~bc}:=\omega_\mu^{~bc}\Varepsilon^\mu_a$.
On a curved manifold whose Riemannian curvature tensor has non-zero components, $\partial_\bcdot\omega_\bcdot^{~\bcdots}\neq0$ is maintained. 
Whereas $\partial_\bcdot\omega_\bcdot^{~\bcdots}$ can be zero at any point in completely flat $\M_4$, an appropriate frame can be found in any $\M_4$, where $\partial_\bcdot\omega_\bcdot^{~\bcdots}\neq0$ is given.
This process is always possible because the spin connection is not a Lorentz tensor in $\TM_4$; thus, it depends on the choice of a local coordinate.
In a completely flat Minkowski manifold, $\partial_\bcdot\omega_\bcdot^{~\bcdots}=1+\sin{\theta}+\cos{\theta}$ in polar coordinate $ds^2=dt^2+dr^2+r^2\hspace{.1em}d\theta^2+r^2\sin^2\hspace{-.2em}{\theta}\hspace{.1em}d\phi^2$.
\noindent
Consequently, non-degenerate $\WWWgr$ always exists in $\Mgr=\varpi\otimes\sigma$.
\vskip .2cm

In summary, classical general relativity has a symplectic structure with respect to the fundamental forms of $d\www^{~ab}\in\Omega^2(\TM)$ and $d\SSS\in\Omega^3(\TM)$. 
These fundamental forms have a coordinate-free cotangent representation of $\WWWgr=d\SSS_\bcdots\wedge d\www^\bcdots/2$ in foliation $\M_5$. 
The pair $\left(\Mgr,\WWWgr\right)$ is proven to be a symplectic manifold.
\QED
%
%
\section{Prequantization}
Geometrical quantization of a symplectic manifold involves two steps\cite{bates1997lectures, nair2005quantum}. The first step is to determine a polarization of a symplectic manifold and to construct Hilbert space $\HH$.
The second step is to introduce automorphism Aut$(\HH,\hat{H})$, which is called a quantum operator and is not treated in this study.
The first step is referred to as prequantization. 
Quantum operators are not defined in a $2n$-dimensional symplectic manifold but are defined in an $n$-dimensional Legendre submanifold.
The step that divides a phase space into half is called polarization.
The Legendre submanifold and polarization are discussed in this section.
The energy eigenvalues depend on the topology of manifold $M$.
For instance, when a Legendre submanifold is a compact space, these operators have discrete eigenvalues.
This topological aspect is realized using a prequantization bundle, which is introduced at the end of this section.

\subsection{Contact manifold}
Symplectic form $\left(\Mgr, \WWWgr\right)$ provides a non-trivial structure of classical general relativity.
Here, we show that the Einstein--Hilbert gravitational Lagrangian can be recognized as a contact form in the context of symplectic geometry.
\begin{definition}\label{CF}{\bf (contact form and contact manifold)}\\
When one-form $\sc$ in $(2n\hspace{-.2em}+\hspace{-.2em}1)$-dimensional manifold $M\otimes\R$ does not vanish at any point of $M$ and two-form $d\sc$ on $\mathrm{Ker}[\sc]$ is non-degenerate, form $\sc$ is called a contact form.
Further, $\left(M\otimes\R, \sc\right)$ is called a contact manifold.
\end{definition}
\noindent
On foliation $\M_5=\bigcap_{\rho\in\R}\varphi_\rho(\M_4)$ defined in Section \ref{nondege}, an extended phase space $(\www,\SSS,\HHH)$ is introduced. 
Scalar Hamiltonian function $H(x)\in\R$ is introduced in $\M_4$ as $\HHH=H(x)\vvv$, and it is added as the third item of the phase space variable as $\Phi:=(\omega^{~a_1a_2}_\mu, S_{b_1b_2;\nu_1\nu_2}, H)$.
\begin{remark}{\bf (contact form for general relativity)}\\
In a manifold $\M_5=\M_4\otimes\R$, four-form
\begin{eqnarray}
\scgr&:=&\frac{1}{2}\SSS_\bcdots\wedge d\www^\bcdots-\HHH
=\LLL\in\Omega^4(\TsM_5)\label{contform}
\end{eqnarray}
is a contact form and $\left(\M_4\otimes\R, \scgr\right)$ is a contact manifold.
Map $\Phi_c:\Omega^4(\TsM_5)\rightarrow\Omega^4(\TsM_5):\sssgr\mapsto\scgr$ is a canonical transformation.
\end{remark}
\begin{proof}
The existence and non-degeneracy follow from those in $\WWWgr$ as presented in Section \ref{2.1}.
Hamiltonian form $\HHH$ also exists, and it is non-degenerate owing to $H\in\varpi\otimes\sigma$ from its definition.

Map $\Phi_c:\Omega^4(\TsM_5)\rightarrow\Omega^4(\TsM_5):\sssgr\mapsto\scgr$ does not change symplectic form $\WWWgr$, i.e.
\begin{eqnarray*}
\Phi_c\WWWgr&=&\Phi_c d(\sssgr)=d(\scgr)=d\left(\SSS_\bcdots\wedge d\www^\bcdots-\HHH\right)
=d\SSS_\bcdots\wedge d\www^\bcdots=\WWWgr,
\end{eqnarray*}
where $d\HHH=d(H\vvv)=\left(\partial_\rho H\right)d\rho\wedge\vvv=0$ because $\partial_\rho H=0$. 
Therefore, map $\Phi_c$ is a canonical transformation.
\end{proof}
\noindent

\subsection{Legendre submanifold}
The Legendre submanifold (foliation) plays an important role in the geometrical quantization of a symplectic manifold, and it is defined as follows:
\begin{definition}\label{lmf}{\bf (Legendre submanifold and foliation)}\\
Suppose that $(M\otimes\R,\sc)$ is a contact manifold. 
When manifold $\LL$ is a submanifold of $M$ with dimension $\mathrm{dim}(\LL)=\frac{1}{2}\mathrm{dim}(M)$ yielding $\sc\bigl|_\LLvac=0$, $\LL$ is called a Legendre submanifold.

If there exists a one-parameter family of Legendre manifolds $\LL=\{\LL_t|t\in\R\}$ that satisfies $\M=\bigcup_t\LL_t$ and $\LL_t\cap\LL_s=0$ for $t\neq s$, $\LL$ is called a Legendre foliation of $\M$.
\end{definition}
Similarly, a Lagrangian submanifold and foliation are defined for symplectic manifold $\left(M,\WWW\right)$ according to the conditions of $\mathrm{dim}(\LL)=\mathrm{dim}(M)/2$ and $\WWW\bigl|_\LL=0$.
A Legendre submanifold for general relativity can be obtained using vacuum solutions of the Einstein equation as follows:
\begin{remark}
The space of vacuum solutions of the classical Einstein equation,
\begin{eqnarray*}
\LLvac:=\left\{
\left(\wwwvac,\eeevac\right)|\epsilon_{a\bcdot\bcdots}
\RRRvac^\bcdots\wedge\eeevac^\bcdot=0=
\TTTvac^a
\right\},
\end{eqnarray*}
provides a Legendre submanifold of contact manifold $\left(\M_5,\scgr\right)$, where
\begin{eqnarray*}
\RRRvac^{ab}:=\RRR^{ab}(d\wwwvac,\wwwvac),&~&
\TTTvac^{a}:=\TTT^{a}(\eeevac,\wwwvac).
\end{eqnarray*}
The dimension of $\LLvac$ is given as $\mathrm{dim}(\LLvac)=10=\mathrm{dim}(\Mgr)/2$.
\end{remark}
\begin{proof}
As mentioned in Section \ref{HamiltonJacobi}, the solution of the classical Einstein equation in vacuum makes the Lagrangian form zero; thus, $\scgr\bigl|_\LLvac=0$ is maintained owing to its definition \ref{contform}.

A vacuum solution of the vierbein and spin forms can be obtained uniquely up to $GL(4)$ symmetry by solving the Einstein equation and a torsionless condition simultaneously; thus, forms $\eee$ and $\www$ are not independent of each other.
When one of the elements is selected from $\LLvac$, for instance $\eeevac$, global Riemannian manifold $(\MM_4\bigl|_\LLvac,\bm{g}_{vac})$ can be fixed with a metric tensor $\left[\bm{g}_{vac}\right]^{\mu\nu}=\eta^{a_1a_2}\left[\Varepsilonvac^{-1}\right]^\mu_{a_1}\left[\Varepsilonvac^{-1}\right]^\nu_{a_2}$.
In addition, at any point in the Riemannian manifold, local Poincar\'{e} manifold $(\M\bigl|_\LLvac,{\bm \eta})$ is associated.
All vacuum solutions $\Varepsilonvac\in\LLvac$ are assumed to be one-to-one parameterized by real parameter space $P$, which induces two bundles as $\Pi_\LL:P\rightarrow\LLvac$ and $\Pi_\M:P\rightarrow\MMvac$.
A lift of $\Pi_\LL^{-1}$ to $\Mvac$ yields homomorphism $\LLvac\simeq\MMvac$, where $\MMvac=\M_4\bigl|_\LLvac$. 
The set of spin forms in Lagrangian submanifold $\MMvac$ is recognized as real polarization, because $\mathrm{dim}(\varpi)=\mathrm{dim}(\sigma)=10$.
In conclusion, $\left(\LLvac,\scvac\right)$ is a Legendre submanifold of contact manifold $\left(\M_5=\M_4\otimes\R,\scgr\right)$, where $\scvac:=\scgr\bigl|_\LLvac$.
\end{proof}
\noindent
For example, a manifold with Schwarzschild solutions of the vacuum equation, which is denoted as $\LLschw$, is a subset of Legendre submanifold $\LLvac\supset\LLschw$.
\begin{example}\label{3.4}
Manifold $\M_4\bigl|_\LLvac$ forms a Legendre foliation of $\M_5=\M_4\otimes[0,\infty)$. 
The flat Lorentz manifold consists of a zero section of $\M_5$.
\end{example}
\vspace{-2mm}
\noindent
The Schwarzschild solution can be uniquely parameterized using a single real number $0\leq m\in\R$, where $m$ is the total mass measured by an asymptotic observer.
When parameter $m$ is considered as the fifth coordinate to immerse  $\M_4$ into $\M_5$, all possible four-dimensional space-time manifolds with Schwarzschild solutions are immersed by mapping $\varphi_{m}$ introduced in Section \ref{nondege}.
A Legendre manifold restricted on the Schwarzschild solution with a fixed $m$ is denoted as $\LLschw(m)$.
This five-dimensional manifold is expressed as $\M_5^{(m)}:=\varphi_m\M_4\bigl|_{\LLschw(m)}$.
Reciprocal map $\Pi_{m}:=\varphi_m^{-1}$ yields bundle $\Pi_{m}:\M_5^{(m)}\rightarrow\M_4\bigl|_{\LLschw(m)}$, and it induces foliation of $\M_5$ with respect to a line bundle with $m\in[0,\infty)$.
We note that a Schwarzschild solution with $m=0$ yields a completely flat Lorentz manifold; thus, the flat Lorentz manifold is a zero section of $\Pi_{m}$.
\QED

\subsection{Prequantization bundle}
A prequantization bundle is defined as follows\cite{doi:10.1142/5731, doi:10.1142/7816}:
\begin{definition}\label{pqb}{\bf (Prequantization bundle)}\\
Suppose that $(M,\WWW)$ is a symplectic manifold such that $\WWW$ is an integral form.
Moreover, it is assumed that $M$ has a principal $G$-bundle with line bundle $\Pi_G:P\rightarrow M$ and structural group $G$.
A connection of the principal bundle is denoted as $\www_G$. 
When a symplectic form is given as $\WWW=d\www_G$, $(\Pi_G,\FFF_G)$ is called a prequantization bundle, where $\FFF_G$ is a curvature associated with connection $\www_G$.
\end{definition}
\begin{example}A $U(1)$-prequantization bundle\end{example}
\vspace{-2mm}
\noindent
A simple example of a prequantization bundle with a principal $U(1)$ group is given here.
Suppose that $(M,\WWW=d\sss)$ is a symplectic manifold with dim$(M)=2n$.   
A principal $U(1)$ bundle is complex-valued line-bundle $\Pi_U:\C\rightarrow M$, and its connection $\www_U$ is introduced in $M$. 
Liouville form $\sss$ induces contact form $\sc_U:=\Pi_U^{~\sharp}\sss$ and contact manifold $(M\otimes\R,\sc_U)$. 
Contact form $\sc_U$ is a $U(1)$ Lie-algebra-valued form and is considered as a differential one-form in $T^*\hspace{-.2em}M$ owing to homomorphism $\R\simeq\uuu(1)$, which is induced by $t\mapsto 2\pi i\hspace{.1em}t$.
When the curvature of the $U(1)$ bundle is given using a contact form such that $d\www_U=\sc_U$ and $\FFF_U:=2\pi i\hspace{.1em}d\www_U$, the cohomology of $\sc_U$ has an integer-valued characteristic class owing to the Chern--Weil theory.
More precisely, $[\sc_U]_\mathrm{dR}\in H^2(M\bigl|_\LL,\R)\subset\mathrm{Im}\hspace{.1em}H^2(M\bigl|_\LL,\Z)$ is obtained, where $\LL$ is a Legendre submanifold.
For instance, the first Chern class is given as $\mathrm{Tr}[\sc_U]=c_1(\FFF_U)$.
Therefore, symplectic manifold $\WWW$ is an integral form; thus, $(\Pi_U, \FFF_U)$ is a prequantization bundle.

Moreover, a symplectic manifold induces a Hamiltonian system in general, and solutions of canonical equations associated with the Hamiltonian system induce Legendre submanifold $\LL$ as an $n$-dimensional hypersurface in $2n$-dimensional manifold $M$.
Therefore, solutions of canonical equations may have topological invariance, e.g. a magnetic monopole in Dirac quantization\cite{Dirac:1931kp} (see also Refs.\cite{frankel_2011,nair2005quantum}).
\QED
\vspace{2mm}

From the example presented above, a prequantization bundle is introduced in general relativity as follows:
Legendre submanifold $(\LLvac\hspace{-.2em}\simeq\hspace{-.2em}\M_5,\WWWgr\hspace{-.2em}=\hspace{-.2em}d\scgr)$ and co-Poincar\'{e} principal bundle in $\M_5$ are considered.
The principal bundle has connection $\AAA_\cP$ and curvature $\FFF_\cP$ as shown in (\ref{cPB1}) and (\ref{cPB2}), respectively.
These forms are introduced in $\M_4$, and they are then immersed in $\M_5$, as discussed in Section \ref{nondege}.
Projection map 
\begin{eqnarray*}
\Pi_4:\left\{\Theta\right\}\rightarrow\left\{\M_4\right\}:\Theta=\{\bm\www,\bm\SSS\}\mapsto\M_4
\end{eqnarray*}
is introduced, where $\left\{\Theta\right\}$ is a set of all possible $\Theta=\{\bm\www,\bm\SSS\}$, and $\left\{\M_4\right\}$ is a set of four-dimensional manifolds with $SO(1,3)$-symmetry. 
This bundle induces a bundle on a base manifold of $\M_5$ such that 
\begin{eqnarray*}
\Pi_5:=\varphi_\rho\circ\Pi_4:\R\otimes\left\{\Theta\right\}\rightarrow\left\{\M_5\right\},
\end{eqnarray*}
where $\left\{\M_5\right\}$ is a set of all possible manifolds introduced in Section \ref{nondege}.
Map $\Pi_5(\Theta|_{\LLvac})=\M_5\bigl|_\LLvac$ induces homomorphism $\LLvac\hspace{-.2em}\simeq\hspace{-.2em}\M_5$.
Contact form $\scgr$ is proven to be divisible modulo $\R/\Z$ owing to {\bf Remark \ref{rem2.1}}:
\begin{eqnarray}
\int_{\M_4}\varphi_\rho^\sharp\scgr&=&8\pi^2\int_{\M_5}c_2(\FFF_\cP),\label{schc}
\end{eqnarray}
with
\begin{eqnarray*}
c_2(\FFF_\cP)\in H^4(\M_4\bigl|_\LLvac\hspace{-.1em}\otimes\R,\R)
\subset\mathrm{Im}\hspace{.1em}H^4(\M_4\bigl|_\LLvac\hspace{-.1em}\otimes\R,\Z).
\end{eqnarray*}
Consequently, the following remark is proven according to the discussion presented above:
\begin{remark}
A prequantization bundle of general relativity in vacuum is given by $\left(\M_5\bigl|_\LLvac,\scvac\right)$.
\end{remark}
\noindent
A Hilbert space for $\LLvac$ was presented in Section \ref{EofMgr}; thus, prequantization is completed.

%
%
\section{EBK quantization}\label{EBKQ}
The last step of geometrical quantization is to construct Hermitian operators on a space $C^\infty(M)$, which is not considered in this study.
Instead, an EBK quantization condition is introduced using a prequantization bundle.
EBK quantization is an extension of Bohr--Sommerfeld quantization, and it provides energy spectra of bounded systems even for a non-variable-separation type of problem.
EBK quantization is applied to a Schwarzschild black hole in this section.
The fundamental physical constants ($G$ and $\hbar$) are explicitly written again hereafter.

%
%
\subsection{The KMA index in general relativity}
First, a simple example of EBK quantization for a system of $N$ classical particles with Hamiltonian $H(q_i,p_i)$, $(i=1,\cdots,N)$, is considered.
The Hamiltonian is assumed to be completely integrable. 
In the case of separation of variables, according to Bohr--Sommerfeld quantization, the Maupertuis action of the symplectic manifold $\left(\R^{2N},\WWW\hspace{-.2em}=\hspace{-.2em}d\sss_i\hspace{-.2em}=\hspace{-.2em}d(p_i\wedge dq_i)\right)$ is expressed as
\begin{eqnarray}
S^\mathrm{BS}_k&=&\frac{1}{2\pi\hbar}\oint_{\Gamma_k}\sss_k=n_k,~~k=1,\cdots,N,\label{EBKc0}
\end{eqnarray}
where $0\leq n_k\in\Z$ is the quantum number of the $k$th variable.
Contour integrations are separately performed along classical closed orbit $\Gamma_k$ of the $k$th particle.
When the phase space variables are not separable, no particle makes a closed orbit; thus, integration (\ref{EBKc0}) of each particle is impossible.
For such a case, Einstein and Brillouin extended the quantization condition (see, for example, Ref.\cite{2004AmJPh..72.1521C}) such that
\begin{eqnarray}
S^\mathrm{EB}_k&=&\frac{1}{2\pi\hbar}\oint_{\Gamma_k}\sum_{i=1}^N\sss_i
=n_k,~~k=1,\cdots,N.\label{EBKc2}
\end{eqnarray}
Contour integrations are performed along all homotopy-independent closed circles on Lagrangian or Legendre submanifold $\LL$.
The contour is also denoted as $\Gamma_k$, and it is not necessarily a classical closed orbit. 
The Bohr--Sommerfeld and Einstein--Brillouin quantization conditions can yield a line spectrum of a hydrogen atom.
However, they cannot reproduce an energy spectrum itself for a harmonic oscillator because it has zero-point energy, which is not included in the quantization conditions.

In 1958, Keller improved\cite{KELLER1958180,KELLER196024} upon this aspect by considering the multivaluedness of a wave function and introduced an additional term to a quantization condition as follows:
\begin{eqnarray}
S^\mathrm{EBK}_k&=&\frac{1}{2\pi\hbar}\oint_{\Gamma_k}\sum_{i=1}^N\sss_i=
n_k+\frac{\mu_k}{4},~~k=1,\cdots,N,\label{EBKc1}
\end{eqnarray}
where $0\leq\mu_i\in\Z$ is an index of the $k$th variable fixed because of the boundary conditions.
This index was first introduced by Keller and then refined by Maslov\cite{maslov75theory,maslov1965wkb} and Arnol'd\cite{Arnol'd1967}.
Therefore, index $\mu_k$ is referred to as the Keller--Maslov--Arnol'd (KMA) index in this study. 
In the case of separation of variables, the KMA index can be obtained as $\mu_k=t_k+2r_k$, where $t_k$ is the number of classical turning points and $r_k$ is the number of reflections with an infinite potential along the trajectory of the $k$th particle.
Later, Maslov and Arnol'd clarified that the index can be recognized as a topological invariant of a double covering space on a Lagrangian submanifold.
The double covering space is introduced to map a dual-valued function on a Lagrangian submanifold to a single-valued one.

For example, for a harmonic oscillator, momentum $p(q)$ as a function of position $q$ is dual-valued as $p(q)=\pm\sqrt{E^2-q^2}$ when the system has total energy $E$.
The trajectory of a particle has two classical turning points; thus, the KMA index is $\mu=t=2$.
Consequently, the energy spectrum has a correct zero-point energy.
Meanwhile, the double covering space of a Lagrangian submanifold for this system is $T^2=S^1\otimes S^1$, which is a simple example of the Arnol'd--Liouville theorem\cite{vogtmann1997mathematical}.
The KMA index of Lagrangian submanifold $\LL$ is defined as $\mu:=H^1(\LL,\Z)$\cite{Arnol'd1967}; thus, for a harmonic oscillator, it can be obtained as $\mu=H^1\hspace{-.1em}\left(T^2,\Z\right)=2$.
This result is consistent with that of Keller's original definition\cite{2004AmJPh..72.1521C}.

The relation between the KMA index and the Chern--Weil theory is discussed in Refs.\cite{Turaev1984, Turaev1987, PACINI2019129}.
The KMA index is defined in the Chern--Weil theory as follows:
Given bundle pair $(E,L)\rightarrow\Sigma$, where $E$ is a symplectic vector bundle over $\Sigma$ and $L$ is a Lagrangian subbundle over $\partial\Sigma$, the KMA index is defined such that (Definition 2.8 in Ref.\cite{10.4310})
\begin{eqnarray}
\mu_\mathrm{CW}(E,L)&:=&\frac{i}{\pi}\int_{\Sigma}\Tr[\FFF_\www]
=2\int_{\Sigma}c_1(\FFF_\www),\label{muCW}
\end{eqnarray}
where $\FFF_\www$ is a curvature with respect to connection $\www$ on $E$, which restricts the boundary of $\Sigma$ to an $L$-orthogonal unitary connection.
It is proven that the KMA index defined by (\ref{muCW}) is equivalent to the standard KMA definition (Theorem 3.1 in Ref.\cite{10.4310}).
In analogy with the definition $\ref{muCW}$, the KMA index for general relativity is defined as follows:
\begin{definition}\label{KMAGR}{\bf (KMA index for general relativity)}\\
The KMA index for general relativity is defined on prequantization bundle $\left(\M_5\bigl|_\LLvac,\scvac\right)$ such that
\begin{eqnarray}
\mugr&:=&\frac{\nugr}{\hbar}
\int_{\LLvac}\varphi_\rho^\sharp\scvac
=\nugr\frac{8\pi^2}{\hbar\kappa}\int_{\M_5|_\LLvac}
c_2\left(\FFF_\cP\bigl|_\LLvac\right),
\end{eqnarray}
where $\nugr$ is an unknown factor corresponding to $1/4$ in ${\mathrm(\ref{EBKc1})}$.
\end{definition}
\noindent
The Planck constant $\hbar$ is used to make the KMA index have null physical dimension.
The exact value of $\nugr$ can be clarified only after the exact quantum general relativity is established

%
%
\subsection{EBK quantization of the Schwarzschild black hole}
We propose an EBK quantization condition of symplectic manifold $(M,\WWW\hspace{-.2em}=\hspace{-.2em}d\sss)$ with prequantization bundle $(\Pi,\sc)$ as follows:
\begin{Pstat}\label{EBKq}{\bf (EBK quantization)}\\
When the Maupertuis action of symplectic manifold $(M,\WWW\hspace{-.2em}=\hspace{-.2em}d\sss)$ has prequantization bundle $(\Pi,\sc)$ and an integral multiple 
\begin{eqnarray}
S^\mathrm{EBK}&=&\frac{1}{\hbar}\int_{\LL}\sss=n,
\end{eqnarray}
the symplectic manifold is EBK-quantized, where $\LL$ is a Legendre submanifold of $M$ with respect to a given Hamiltonian form and $0\leq n\in\Z$ is called the quantum number.
\end{Pstat}
\noindent
We note that a Hamiltonian (and thus symplectic and contact forms) has null physical dimension in this study.
Hereafter, pull-back mapping is omitted in the expressions for simplicity.
Immersion of $\M_4$ into $\M_5$ is performed as defined in Section \ref{nondege}. 
The KMA index is not explicitly introduced in the definition; it naturally appears through the prequantization bundle.

When {\bf physical statement \ref{EBKq}} is applied to the vacuum solutions of the Einstein equation, the following quantization condition can be obtained:
\begin{Pstat}\label{EBKvac}{\bf (vacuum solution)}\\
The Maupertuis action for a vacuum solution of the Einstein equation can be expressed as
\begin{eqnarray}
\frac{1}{\hbar}\int_{\LLvac}\sssvac&=&\int_{\LLvac}
\left(\frac{1}{\hbar}\scvac+\HHH\right)=n.
\end{eqnarray}
Integration of the vacuum contact form yields the second Chern class; thus, a quantization condition can be obtained as
\begin{eqnarray}
{E}_{{\hspace{-.1em}vac}}&=&n+\frac{\mugr}{\nugr},\label{Evac}
\end{eqnarray}
where the KMA index appears through {\bf Definition \ref{KMAGR}}.
\end{Pstat}
\noindent
Here, vacuum energy ${E}_{{\hspace{-.1em}vac}}$ has null physical dimension.
The KMA index of general relativity $\mugr$ can be evaluated using the cohomology of a Lagrangian submanifold.
Ambiguity still exists in choosing a value of ${\nugr}$.
We note that (\ref{contform}) must be $\sc=\frac{1}{2\kappa\hbar}\SSS_\bcdots\wedge d\www^\bcdots-\HHH$ after introducing the fundamental constants back into the expression.
The KMA index appears as the second Chern class of the contact form.

The Schwarzschild solution of the classical Einstein equation is given with an asymptotically flat polar coordinate $x^a=(t,r,\theta,\phi)$ as follows:
\begin{eqnarray}
\eee^a&=&\left(
f~dt,
f^{-1}~dr,
r~d\theta,r\sin{\theta}~d\phi\right),\label{virschw}
\end{eqnarray}
and
\begin{eqnarray}
\vomega^{ab}=\left(
\begin{array}{cccc}
0&-G\hspace{.1em}m_{\hspace{-.1em}S}/r^2~dt~& 0 & 0 \\
~& 0 &  f~d\theta &
f\sin{\theta}~d\phi \\
~&~&0&~~\cos{\theta}~d\phi\\
~&~&~&0\\
\end{array}
\right),\label{www}
\end{eqnarray}
where $f^2(r)=1-2m_{\hspace{-.1em}S}G/r$ and $m_{\hspace{-.1em}S}$ is the mass of the black hole measured by an asymptotic observer.
The classical Hamiltonian can be obtained as
\begin{eqnarray}
\HHH_{\mathrm{Schw}}&=&-\frac{1}{2\kappa\hbar}\vomega^{\bcdot}_{~\star}\wedge\vomega^{\star\bcdot}
\wedge\SSS_\bcdots
=-\frac{1}{\kappa\hbar}\sin{\theta}~dt\wedge dr\wedge d\theta\wedge d\phi.\label{Hscw}
\end{eqnarray}
We note that the Hamiltonian form of a Schwarzschild solution does not include the black-hole mass as expressed in (\ref{Hscw}); thus, the assumption $\partial H/\partial m_{\hspace{-.1em}S}=0$ is also true in this case.
The total energy in a sphere with radius $R$ can be obtained by integrating (\ref{Hscw})  as
\begin{eqnarray}
E_{\mathrm{Schw}}(R)&:=&
\int^{t_p}\int^R\int^{S^2}\HHH_{\mathrm{Schw}}=-\frac{R}{l_p}.\label{Eschw}
\end{eqnarray}
The energy density per unit surface area of the sphere is expressed as
\begin{eqnarray*}
\overline{E}_{\mathrm{Schw}}(R)&:=&\frac{E_{\mathrm{Schw}}}{4\pi (R/l_p)^2}=-\frac{1}{4\pi}\frac{1}{(R/l_p)};
\end{eqnarray*}
thus, the energy density at the event horizon of a black hole measured by an asymptotic observer can be obtained as
\begin{eqnarray*}
m_p\Bigl|\overline{E}_{\mathrm{Schw}}(R=2m_{\hspace{-.1em}S} G)\Bigr|&=&
\frac{\hbar}{8\pi G\hspace{.1em}m_{\hspace{-.1em}S}}.
\end{eqnarray*}
We note that the energy density of space-time in a given region is a coordinate-dependent observable.
The energy density given here is the value measured by the asymptotic observer.
This energy density is none other than the Hawking temperature\cite{hawking1975}.

Quantization condition (\ref{Evac}) is applied to the energy of a black hole, (\ref{Eschw}), as
\begin{eqnarray*}
|E_{\mathrm{Schw}}(R=2m_{\hspace{-.1em}S}G)|&=&
2\frac{m_{\hspace{-.1em}S}}{m_p}=n+\frac{\mugr}{\nugr},
\end{eqnarray*}
Consequently, the mass spectrum of a Schwarzschild black hole can be obtained as
\begin{eqnarray}
m_{\hspace{-.1em}S}&=&\frac{m_p}{2}\left[n+\frac{\mugr}{\nugr}\right].\label{Mschw}
\end{eqnarray}
Integer index $\mugr$ can be obtained in $\mathrm{Im}\hspace{.1em}H^4(\M_5,\Z)$ according to the topology of a four-dimensional Schwarzschild black hole immersed in five-dimensional space.
We note that the Schwarzschild solution is a single-valued function; thus, it does not require a covering space. 
A simple vacuum solution of the Schwartzschild black hole at the center of the universe is considered here.
Five-dimensional manifold $\M_5=\R\otimes\M_4$ is a Legendre foliation introduced in {\bf Remark \ref{lmf}}.
Given that a Schwarzschild solution has a singularity at the three-dimensional center of a black hole, this point must be removed from the integration region.
In addition, a periodic condition is required in the time coordinate in (\ref{virschw}) because the Schwarzschild solution is static.
Therefore, the integration manifold is homeomorphic as $\M_5=\R\otimes T^1\otimes(\R^3\setminus\{\bm{0}\})$.
The origin of the fifth coordinate indicates a flat Minkowski space; thus, no singularity exists.
The fifth coordinate $m_{\hspace{-.1em}S}=[0,\infty)$ is deformation-retractable to a point when the one-point compactification is applied by adding  ${m_{\hspace{-.1em}S}=\infty}$; hence, the cohomology becomes $H^4(\M_5,\Z)\cong H^4(T^1\otimes(\R^3\setminus\{\bm{0}\}),\Z)$.
Given that the upper bound of an $r$-integration is $R$ (black-hole radius), the topology of the integration region in a spatial three-dimensional manifold contains $\R^3\setminus\{\bm{0}\}\cong S^2\otimes[R_0,R]$, where $0<R_0<R$.
Therefore, cohomology $H^4(T^1\otimes S^2\otimes[R_0,R],\Z)\cong H^0(T^1\otimes S^2\otimes[R_0,R],\Z)=1$ is obtained because $T^1\otimes S^2\otimes[R_0,R]$ is a connected and compact manifold.

In conclusion, the mass spectrum of the Schwarzschild black hole is fixed as 
\begin{eqnarray}
m_{\hspace{-.1em}S}&=&\frac{m_p}{2}\left(n+\frac{1}{\nugr}\right),~~n=0,1,2,\cdots,\label{Mschw0}
\end{eqnarray}
up to an ambiguity of ${\nugr}$.

\section{Summary and discussion}
A symplectic manifold of general relativity for a pure gravitational case was presented as {\bf Theorem \ref{sypform}}.
Fundamental forms of the symplectic manifold were obtained from the principal bundle with co-Poincar\`{e} symmetry as a structural group.
Further, according to the standard procedure of geometrical quantization, a prequantization bundle of general relativity was constructed.
The key to quantization is that the contact form should have a topological class (second Chern class), which was proved by the current author as {\bf Theorem 4.2} in Ref.\cite{doi:10.1063/1.4990708}.
On the basis of the symplectic manifold of general relativity, EBK quantization was constructed with the second Chern class as the KMA index.

When the vierbein form was given as a vacuum solution of the Einstein equation, the surface and spin forms were uniquely determined under a torsionless condition.
Simultaneously, a Riemannian geometrical structure of manifold $\M_4$ was completely fixed by a metric tensor given by the vierbein form.
These mappings induced homomorphisms as shown in Table \ref{tbl1}.
\begin{table}[t]
\begin{center}
\begin{tabular}{ccccccc}
$\Varepsilonvac$&$\rightarrow$&
$\LLvac$&$\rightarrow$&
$\SSS\bigl|_{\LLvac}$
\\
~&~&\rotatebox{90}{$\simeq$}&$\searrow$&\rotatebox{90}{$\simeq$}
\\
~&~&$\M_4\bigl|_\LLvac$&~&$\www\bigl|_{\LLvac}$
\end{tabular}
\caption{Homomorphism network of vacuum solutions of the Einstein equation in vacuum.}\label{tbl1}
\end{center}
\end{table}
Therefore, the KMA index of the topological invariance can provide a constraint on the structure of the space-time manifold itself.
By contrast, e.g. the standard EBK quantization of a system of mass points provides a constraint on the phase space, which consists of the position and momentum of particles.


As an application of EBK quantization to general relativity, we discussed the mass spectrum of a Schwarzschild black hole.
In general, a connection bundle is called a ``flat bundle'' when it provides zero curvature at any point in a manifold.
When its Riemannian curvature is also zero, it is referred to as a ``trivial flat bundle''.
A globally flat Lorentz manifold is an example of a trivial flat bundle, and its characteristic classes are trivial.
We introduced a flat bundle in general relativity as a bundle with zero Ricci curvature at any point in a manifold.
When the flat Lorentz manifold was immersed in $\M_5=\M_4\left(\LLschw\right)\otimes\R$ as a zero section of the Schwarzschild foliation such as that in {\bf Remark \ref{3.4}}, it was able to contain a non-trivial characteristic class and showed a rich structure.
In reality, the vacuum solutions of the Einstein equation include a complicated structure, such as black hole solutions. 
The investigation of the gravitational vacua of classical general relativity was reduced to a mathematical investigation of a flat bundle in a space-time manifold.

Quantization of a black hole has a long history.
First, Bekenstein discussed the mass spectrum of a Kerr black hole as an analogy of a charged particle with spin\cite{Bekenstein1974}.
Louko and Winters-Hilt extended the discussion to Reissner--Nordstr$\ddot{\rm o}$m--anti-de Sitter black holes using Hamiltonian thermodynamics\cite{PhysRevD.54.2647}.
An area spectrum of black holes has also been discussed on the basis of a quantum area operator by many authors\cite{0264-9381-14-1A-006,0264-9381-15-6-001,0264-9381-18-22-310,0264-9381-20-9-305,0264-9381-25-20-205014}.
Other approaches from string theory\cite{Kiselev:2004vx} and loop quantum gravity\cite{PhysRevLett.110.211301} have also been investigated.  
An independent result of a black-hole mass spectrum based on EBK-quantized general relativity was given in the present study.
The mass spectra obtained in previous studies\cite{Bekenstein1974,0264-9381-20-9-305,0264-9381-18-22-310,Bekenstein:1997bt} are apparently different from those of our result.
While the mass spectrum of black holes was obtained as a consequence of an area spectrum (including effects from charge and angular momentum) in previous studies, the spectrum condition was set on the mass square rather than the mass itself.
For example, the mass spectrum of the Schwarzschild black hole was given as
\begin{eqnarray*}
\widetilde{m}_s^2&=&\frac{m^2_{p}}{2}\left[n+\tilde\mu\right],
\end{eqnarray*}
where $\tilde\mu=0$ was proposed by Bekenstein\cite{Bekenstein1974,Bekenstein:1997bt} and Medbed\cite{0264-9381-25-20-205014}, and $\tilde\mu=1/2$ was proposed by Barvinsky, Das, Kunstatter\cite{0264-9381-18-22-310} and Gour, Medved\cite{0264-9381-20-9-305}.
In contrast to previous results, EBK quantization gave a constraint on the black-hole mass itself.
Our result is rather natural when considering a process in which two black holes merge into one.
Suppose that black holes $BH_1$ and $BH_2$, whose masses are $m_1$ and $m_2$, respectively, merge into black hole $BH_3$ whose mass is $m_3$.
The merging process must satisfy the energy conservation law as $m_3=m_1+m_2$.
Furthermore, each mass is represented as $m_i^2={m^2_{p}}\hspace{.1em}n_i/2$ under a constraint on the mass square; thus, the relation $\sqrt{n_3}=\sqrt{n_1}+\sqrt{n_2}$ is obtained.
Here, $\tilde\mu$ is set to zero for simplicity.
Therefore, black-hole merging can occur only in the case of two black holes whose masses satisfy the relation $\sqrt{n_1n_2}\in\N$.
Meanwhile, a mass constraint obtained from EBK quantization does not induce such a limitation.
In contrast, only an isolated black hole is considered in this study.
A superposition of two Schwarzschild black holes at different spatial points is not a solution of the Einstein equation owing to its non-linearity.
The mass spectra discussed here correspond to those of an asymptotic field without any interactions (except self-interactions) in the quantum field theory.
We note that the discussion about the phenomena of a two black-hole merger must be carefully studied before the quantum correction for a highly non-linear gravitational field can be clarified.

The cosmological constant $\Lambda_c$ cannot appear in a topological Lagrangian because the volume form is not co-Poincar\'{e} invariant\cite{doi:10.1063/1.4990708} at a classical level.
An effective cosmological constant may appear at a quantum level as a vacuum fluctuation.
This possibility was discussed by the current author in Ref.\cite{Kurihara:2016gyt}.

EBK quantization (as well as Bohr--Sommerfeld quantization) is an approximation method for choosing quantum-mechanically possible solutions from among the classical ones.
For example, for a hydrogen atom, classical electrodynamics provides a continuous energy spectrum as a solution of a non-relativistic Newtonian equation of motion for an electron in a classical Coulomb electric field.
The Bohr--Sommerfeld quantization condition specifies quantum-mechanically possible orbits from the classical ones and succeeds in explaining the discrete spectrum of a hydrogen atom.
The same results can be obtained as an interaction between the classical fermion and electromagnetic fields.
We note that spinor representation is one of the representations of a $SO(1,3)$ group; thus, the spin of a particle is a classical object and does not require quantization. 
The energy spectrum obtained using the Bohr--Sommerfeld quantization condition remains a classical solution, and the real quantum effect was initially observed as the Lamb shift\cite{PhysRev.72.241}.
A systematic method to calculate the quantum corrections in an electromagnetic interaction was established as quantum electrodynamics, which is now a paragon of field quantization.
We must develop a method for calculating the true quantum effects beyond the EBK approximation.
For instance, the mass spectrum of the Schwarzschild black hole obtained using EBK quantization has an ambiguous parameter ${\nugr}$, whose true value can be clarified only after quantum gravity is established.

Before we attempt to establish a sophisticated quantum gravity such as the quantum field theory, we can learn from the history of the old quantum mechanics.
Although EBK quantization cannot provide any quantum equation or quantum state vector, they can be discussed using geometrical quantization.
Quantum corrections for the classical solutions can be calculated using the Wentzel--Kramers--Brillouin (WKB) method.
The WKB method for general relativity beyond vacuum solutions was discussed  by the current author in Ref.\cite{2017arXiv171207964K}.
The final objective is to establish a calculable method for measurable quantities of the quantum effects on gravity. 
A possible next step for constructing a quantum general relativity was discussed by the current author in Ref.\cite{2017arXiv170305574K}.

\subsection*{Acknowledgment}
I appreciate the kind hospitality of all the members of the theory group of Nikhef, particularly Prof. J. Vermaseren and Prof. E. Laenen.
A major part of this study was conducted during my stay at Nikhef in 2017.
I would also like to thank Dr.~Y.~Sugiyama and Prof. J. Fujimoto for their constant encouragement and insightful discussions.
%
%
\bibliographystyle{elsarticle-num}
\bibliography{ref}
\end{document}